%% file: KernelBruteforce.tex
\documentclass[conference,letterpaper]{IEEEtran}
\usepackage[cmex10]{amsmath}
\usepackage{amsthm,amssymb}
\usepackage[cp1251]{inputenc}
\usepackage{graphicx}
\usepackage{epstopdf}
\usepackage{color}
\usepackage{flushend}
\usepackage{subcaption}
\usepackage{array}
\usepackage{float}
\usepackage[ruled,vlined,linesnumbered]{algorithm2e}
\usepackage{setspace}
\usepackage{multirow}
\usepackage{multicol}

\input{defs}

\begin{document}

\title{A Search Method for Large Polarization Kernels}
\author{
\IEEEauthorblockN{Grigorii Trofimiuk}
\IEEEauthorblockA{ITMO University, Russia\\
Email: gtrofimiuk@itmo.ru}}

\maketitle

\begin{abstract}
A new search method for large polarization kernels is proposed. The algorithm produces a kernel with given partial distances by employing depth-first search combined with some methods for search space reduction. Using the proposed method, we improved almost all existing lower bounds on the maximum rate of polarization for kernels of size from 17 to 27.

We also obtained kernels which admit low complexity processing by the recently proposed recursive trellis algorithm. Numerical results demonstrate the advantage of polar codes with the proposed kernels compared with shortened polar codes and polar codes with small kernels.
\end{abstract}



\input{introduction}

\input{background}
\input{search}

\input{maximization}
\input{complexity}
\input{numeric}
\input{conclusions}
\input{kernels}


\end{document}

%% file: defs.tex
\theoremstyle{plain}
\newtheorem{lemma}{Lemma}
\newtheorem{proposition}{Proposition}
\DeclareMathOperator{\wt}{wt}

\DeclareMathOperator{\bbS}{\mathbb S}

\DeclareMathOperator{\bbM}{\mathbb M}

\renewcommand{\ell}{l}

\newcommand{\set}[1]{\left\{{#1}\right\}}

\newcommand{\mC}{\mathcal{C}}
\newcommand{\mD}{\mathcal{D}}

\newcommand{\mF}{\mathcal{F}}

\newcommand{\mK}{\mathcal{K}}

\newcommand{\bF}{\mathbb{F}}

\newcommand{\bK}{\mathbb{K}}

\newcommand{\bM}{\mathbb{M}}

\newcommand{\bW}{\mathbb{W}}



%% file: introduction.tex
\section{Introduction}
Polar codes are a novel class of error-correcting codes, which achieves the symmetric capacity of a binary-input memoryless channel  $W$. They have low complexity construction, encoding, and decoding algorithms \cite{arikan2009channel}. 

Polarization is a general phenomenon and it is not restricted to the Arikan matrix \cite{korada2010polar}.  One can replace it by a larger matrix, called \textit{polarization kernel}, which has better polarization properties, thus obtaining  better finite length performance. Polar codes with large kernels were shown to provide an asymptotically optimal scaling exponent \cite{fazeli2018binary}.
Some fundamental limitations on the rate of polarization together with various kernel constructions were proposed in \cite{korada2010polar}, \cite{presman2015binary}, \cite{lin2015linear}.

The rate of polarization of the kernel depends only on its \textit{partial distances}. 
In this paper, we propose a search algorithm for polarization kernels with given partial distances. The proposed approach is based on depth-first search combined with some methods for search space reduction. This allows us to find polarization kernels of size $17 \leq l \leq 27$ (except $l=24$) with better rate of polarization compared with the ones constructed in \cite{korada2010polar}, \cite{presman2015binary}.

We also demonstrate the application of the proposed approach to the construction of polarization kernels which admit low complexity processing by the recently proposed recursive trellis algorithm \cite{trifonov2021recursive, trifonov2019trellis}. Simulation results show that polar codes with the obtained large kernels provide significant performance gain compared with shortened polar codes and polar codes with small kernels.

%% file: background.tex
\section{Background}
\label{sBackground}
\subsection{Notations}
For a positive integer $n$, we denote by $[n]$ the set of $n$ integers $\{0,1,\dots ,n-1\}$. The vector $u_a^b$ is a subvector $(u_a,u_{a+1},\dots,u_b)$ of some vector $u$. For vectors $a$ and $b$ we denote their concatenation by $a.b$. $M[i]$ is the $i$-th row of the matrix $M$. By $M[i,j]$ we denote the $j$-th element of $M[i]$. For matrices $A$ and $B$ we denote their Kronecker product by $A \otimes B$. By $A^{\otimes m}$ we denote $m$-fold Kronecker product of matrix $A$\  with itself.
\subsection{Polarizing transformation}
\label{ss:pt}
 A \textit{polarization kernel} $K$ is a binary invertible $l \times l$ matrix, which is not upper-triangular under any column permutation \cite{korada2010polar}.
 An $(n, k)$ mixed-kernel polar code \cite{presman2016mixed}, \cite{bioglio2020multikernel} is a linear block code generated by $k$ rows of matrix $G_n = \mK_1 \otimes \mK_2\otimes ... \otimes \mK_s$, where $\mK_i$ is an $l_i \times l_i$ polarization kernel and $n = \prod_{i=1}^s l_i$.
The encoding  scheme is given by
$ c_0^{n-1}= u_0^{n-1}G_n$,
where $u_i,i\in \mathcal F$ are  set to some pre-defined values, e.g. zero (frozen symbols),  $|\mF| = n - k$, and the remaining values $u_i$ are set to the payload data. 


\subsection{Fundamental parameters of polar codes}
\subsubsection{Rate of polarization}
Let $W: \{0,1\} \to \mathcal{Y}$ be a symmetric binary-input discrete memoryless channel (B-DMC) with capacity $I(W)$. By definition,
$$
I(W) = \sum_{y \in \mathcal Y} \sum_{x \in \set{0,1}} \frac{1}{2}W(y|x) \log \frac{W(y|x)}{\frac{1}{2}W(y|0)+\frac{1}{2}W(y|1)}.
$$

The Bhattacharyya parameter of $W$ is 
$$Z(W) = \sum_{y\in\mathcal{Y}} \sqrt{W(y|0)W(y|1)}.$$

Consider the polarizing transform $K^{\otimes m}$, where $K$ is an $l \times l$ polarization kernel, and bit subchannels $W_{m}^{(i)}(y_0^{n-1},u_0^{i-1}|u_i)$, induced by it. Let $Z_m^{(i)} = Z(W_{m}^{(i)}(y_0^{n-1},u_0^{i-1}|u_i))$ be a Bhattacharyya parameter of $i$-th subchannel, where $i$ is uniformly distributed on the set $[l^m]$. Then, for any B-DMC $W$ with $0<I(W)<1$,
we  say that an $\ell\times\ell$ matrix $K$ has the rate of polarization (also known as error exponent) $E(K)$ if \cite{korada2010polar}
\begin{itemize}
\item[(i)] For any fixed $\beta < E(K)$,
\[
\liminf_{n \to \infty} \Pr[Z_n \leq 2^{-\ell^{n\beta}}] = I(W).
\]
\item[(ii)] For any fixed $\beta > E(K)$,
\[
\liminf_{n\to\infty} \Pr[Z_n \geq 2^{-\ell^{n\beta}}] = 1.
\]
\end{itemize}


Suppose we construct $(n,k)$ polar code $\mathcal C$ with kernel $K$. Let $P_e(n)$ be a block error probability of $\mathcal C$ under transmission over $W$ and decoding by SC algorithm. It was proven \cite{korada2010polar}, that if $n/k <\ I(W)$ and $\beta < E(K)$, then for sufficiently large $n$
the probability $P_e(n)$ can be bounded as 
$
P_e(n) \leq 2^{-n^\beta}.
$ 

It turns out that the rate of polarization is independent of channel $W$. Namely, let $\langle g_1,g_2,\dots,g_k \rangle$ be a linear code, generated by vectors $g_1,g_2,\dots, g_k$. Let $d_H(a,b)$ be the Hamming distance between $a$ and $b$. Let $d_H(b,\mathcal C) = \min_{c\in \mathcal C} d_H(b,c)$ be a minimum distance between vector $b$ and linear block code $\mathcal C$. Let $\wt(b) = d_H(b,\mathbf 0)$. 

The \textit{partial distances}(PD) $\mD_i, i = 0,\dots,l-1$, $l\times l$ of the kernel $K$ are defined as follows:
\begin{align}
\label{fPDDef}
\mD_i &= d_H(K[i],\langle K[i+1],\dots,K[l-1] \rangle), i \in [l-1],\\
\mD_{l-1} &= d_H(K[l-1],\mathbf 0).
\end{align}

The vector $\mD$ is referred to as a \textit{partial distances profile} (PDP). It was shown in \cite{korada2010polar} that for any B-DMC $W$ and any $l\times l$ polarization kernels $K$ with PDP $\mD$, the rate of polarization $E(K)$ is given by 
\begin{align}
E(K) = \frac{1}{l}\sum^{l-1}_{i=0}\log_l \mD_i.
\label{f:Rate}
\end{align}


It is possible to show that by increasing kernel dimension $l$ to infinity, it is possible to obtain $l\times l$ kernels with rate of polarization arbitrarily close to 1  \cite{korada2010polar}. Explicit constructions of kernels with high rate of polarization are provided in \cite{presman2015binary,lin2015linear}.
\subsubsection{Scaling exponent}
Another crucial property of polarization kernels is the \textit{scaling exponent}.
Let us fix a binary discrete memoryless channel $W$ of capacity $I(W)$ and a desired block error probability $P_e$. Given $W$ and $P_e$,
suppose we wish to communicate at rate $I(W) - \Delta$ using a family of $(n,k)$ polar codes with kernel $K$. The value of $n$
scales as $O(\Delta^{- \mu(K)})$, the constant $\mu(K)$ is known as the scaling exponent \cite{hassani2014finitelength}. The scaling exponent depends on the channel. Unfortunately, the algorithm of its computation is only known for the case of the binary erasure channel \cite{hassani2014finitelength, fazeli2014scaling}. It is possible to show \cite{pfister2016nearoptimal,fazeli2021binary} that there exist $l\times l $ kernels $K_l$, such that $\lim_{l\rightarrow \infty}\mu(K_l)=2,$
i.e. the corresponding polar codes provide an optimal scaling behaviour. 
Constructions of the kernels with good scaling exponent are provided in 
\cite{trofimiuk2019construction32,yao2019explicit}.

%% file: search.tex
\section{Search algorithm}
 In this section we describe an algorithm, which searches for an $l\times l$ kernel with the required partial distance profile. We start from a basic idea and add some improvements, which reduce the overall search space.
\subsection{Basic algorithm}
For given $l\times l$ kernel $K$ we define a sequence of 
$(l, l-\phi,d_K^{(\phi)})$ \textit{kernel codes} $\mC_{K}^{(\phi)} = \langle K[\phi],\dots,K[l-1] \rangle$, $\phi \in [l],$ and $\mC_K^{(l)}$ contains only zero codeword. Suppose we want to obtain an $l \times l$ polarization kernel $K$ with given PDP $\mD$. Our idea is to successively construct such a kernel starting from row $K[l-1]$ to row $K[0]$. It is convenient to define sets of \textit{candidate rows} $\mathbb M_\phi, \phi \in [l]$, i.e. for each obtained $K$ its rows $K[\phi] \in \mathbb M_\phi$. The definition of partial distances allows us to initialize these sets as  $\bM_\phi^{(def)} = \set{v_0^{l-1}|v_0^{l-1} \in \bF_2^{l}, \wt(v) \geq \mD_\phi}.$

\begin{algorithm}
\caption{\texttt{BasicKernelSearch}$(K, \phi, \bM, \mD)$}
\label{alg_basic}
\If{$\phi = -1$}{
        \Return $K$;\\
}
\For{\textbf{each} $v \in \bbM_\phi$}{ \label{basicLoop}
        $d \gets d_H(v, \mC_{K}^{(\phi+1)})$;\\ \label{basicPDcomp}
        \If{$d = \mD_\phi$}{
                $K[\phi] \gets v$;\\
                $\widehat K \gets $ \texttt{BasicKernelSearch}$(K, \phi-1, \bM, \mD)$;\\
                \If{$\widehat K \ne \mathbf 0^{l \times l}$}{
                        \Return $\widehat K$ \label{basicLoopReturn};\\
                }
        }
}
\Return{$\mathbf 0^{l \times l}$};\\
\end{algorithm}

This approach can be described in terms of depth-first search over candidate rows, which is illustrated in Alg. \ref{alg_basic}. One should start search by calling \textit{BasicKernelSearch}$(\mathbf 0^{l \times l}, l-1, \bM_\phi^{(def)}, \mD)$, where $\mathbf 0^{l \times l}$ is a zero $l\times l$ matrix. At \textit{phase} $\phi$ the algorithm iterates through 
vectors $v \in \mathbb M_\phi$ and checks whether $d_H(v, \mathcal C_{K}^{(\phi+1)}) = \mD_{\phi}$. If such a row is found, the algorithm sets $K[\phi]\leftarrow v$, proceeds to phase $\phi - 1$ and starts new search over rows $v \in \bbM_{\phi-1}$. If there are no such rows, we return to phase $\phi+1$ and continue iterating through $v \in \bbM_{\phi+1}$. The algorithm traverses the recursion tree until a kernel $K$ with required PDP $\mD$ is obtained or zero matrix $\mathbf 0^{l \times l}$ is returned.

\subsection{Evaluation of partial distances}
Let $A_K^{(\phi)}$ and $A_K^{(\phi)}(v)$ be weight distributions of $\mC_K^{(\phi)}$ and coset $\mC_K^{(\phi)} \oplus v$, respectively. Observe that \begin{equation}
\label{fAA}
A_K^{(\phi-1)} = A_K^{(\phi)}+A_K^{(\phi)}(K[\phi-1]).
\end{equation}

Equation \eqref{fAA} implies that to obtain all partial distances one can compute the weight distribution of all kernel codes. A good survey of weight distribution computation algorithms can be found in \cite{bouyukliev2020characteristic} together with a new method based on characteristic vector representation of linear codes.

In this work, for $\phi \geq l/2$, we compute partial distance directly by definition \eqref{fPDDef}. This allows us to stop the computation of partial distance $d_H(v, \mathcal C_{K}^{(\phi+1)})$ once the row with weight less than $\mD_{\phi}$ is occurred, which quickly prunes unsuitable rows. 
 
Let $\bar \mC_K^{(\phi)}$ be dual code of the kernel code $\mC_K^{(\phi)}$. For phases $\phi < l/2$ equation \eqref{fAA} enables us to compute $d_H(K[\phi-1], \mathcal C_{K}^{(\phi)})$ by calculation of weight distribution of codes $\bar \mC_K^{(\phi)}$, $\bar \mC_K^{(\phi-1)}$,  and application of MacWilliams identities.  

\subsection{Restriction of candidate rows}
\label{ssRestriction}
\begin{proposition}[\cite{fazeli2014scaling}]
\label{pAddRow}
Let $K$ and $\widehat K$ be $l \times l$ polarization kernels, where $\widehat K$ is obtained from $K$ by adding row $i$ to row $j$ with $j <\ i$, then $\mC_{\widehat K}^{(i)} = \mC_{K}^{(i)}$ for all $i \in [l]$. Naturally, $E(\widehat K) = E(K)$ and $\mu(\widehat K) = \mu(K)$.
\end{proposition}
Proposition \ref{pAddRow} implies that any kernel $K$ can be transformed by addition of row $i$  to row $j$ with $j < i$ into kernel $\widehat K$ with $\mD_i = \wt(\widehat K[i])$, $i \in [l],$ and same polarization properties as $K$. Therefore, without loss of generality, the sets of candidate rows can be defined as
\begin{equation}
\label{fCandSet} 
\bbM_\phi^{(r)} = \set{v_0^{l-1}| v_0^{l-1} \in \bF_2^l, \wt(v) = \mD_\phi}.
\end{equation} 

Note that one can further restrict $\bbM_\phi$ to construct more specific kernels, as it was done in \cite{trofimiuk2019construction32}, but in this paper we consider the candidate sets \eqref{fCandSet} only. 

Despite the noticeable restriction \eqref{fCandSet} of $\bbM_\phi$, function \textit{BasicKernelSearch} is still infeasible for large $l$. In the next sections we modify this algorithm to reduce the search space.

\subsection{Syndrome computation}

   Let $H_K^{(\phi)}$ be a parity-check matrix of the kernel code $\mC_K^{(\phi)}$. Observe that if vectors $v$ and $v'$ are from the same coset, i.e. 
$vH_K^{(\phi)} = v'H_K^{(\phi)}$, then, $d_H(v, \mC_K^{(\phi)}) = d_H(v', \mC_K^{(\phi)})$. At each phase $\phi$ we propose to store the set $\bbS_\phi$ of syndromes $vH_K^{(\phi+1)}$ of cosets $\mC_K^{(\phi+1)} \oplus v$, for which the partial distance $d_H(v, \mC_{K}^{(\phi+1)})$ has already been computed.   
 Thus, if for some row $v'$ its syndrome $v'H_K^{(\phi)} \in \bbS_\phi$, then $v'$ can be skipped in the loop at lines \ref{basicLoop}-\ref{basicLoopReturn} of Alg \ref{alg_basic}.


\subsection{Reducing the search space}

Let $\mC_{K,v}^{(\phi)}$ denote the code $\langle v, K[\phi],\dots,K[l-1] \rangle$. We say that $l \times l$ binary polarization kernels $K$ and $\hat K$ are equivalent if there exists an $l \times l$ permutation matrix $P$ such as $\hat K = K P$. This definition implies that the kernel codes $\mC_K^{(i)}$ and $\mC_{\hat K}^{(i)}, i \in [l]$, are also equivalent. We propose to compute $d_H(v', \mC_{K}^{(\phi+1)})$ only if code $\mC_{K,v}^{(\phi+1)}$ is not equivalent to \textit{all other} codes $\mC_{\widehat K,\widehat v}^{(\phi)}$  already arisen at phase $\phi$ of Alg. \ref{alg_basic}.

Unfortunately, to the best of our knowledge, there are no simple algorithms for testing code equivalence. It was shown in \cite{petrank1997code} that code equivalence problem can be reduced to the graph isomorphism problem in polynomial time. The support splitting algorithm \cite{sendrier2000finding}\ can be used to find a permutation between equivalent codes. Code equivalence test based on  canonical augmentation is proposed in \cite{bouyukliev2007about, bouyukliev2020computer}.

The code equivalence problem is hard to decide. Instead of this, we propose to consider the codes $\mC_{K,v}^{(\phi)}$ with unique weight distribution $A_K^{(\phi+1)}(v)$ of the coset $\mC_K^{(\phi+1)} \oplus v$. Observe that this distribution can be directly obtained while computing of the partial distance $d_H(v, \mC_K^{(\phi)})$. Certainly, nonequivalent codes may have the same weight distribution, therefore, if we use this modification, the algorithm is not guaranteed to find a kernel with given PDP even if it exists.

\subsection{Proposed search algorithm}
Alg. \ref{alg_full} presents a proposed search algorithm for large polarization kernels.  One should start search by calling \textit{KernelSearch}$(\mathbf 0^{l \times l}, l-1, \bbM_\phi^{(r)})$. We assume that the variables $\bK$ and $\bW$ are global. We also modified the basic algorithm to collect several polarization kernels  with PDP $\mD$ into the set $\bK$. This set is used in Section \ref{sComplexity}. The set $\bW_\phi$ is a set of weight distributions $A_{K}^{(\phi+1)}(v)$. We used universal hashing \cite{thorup2020high} for efficient comparison of weight distributions.

\begin{algorithm}
\SetNoFillComment
\caption{\texttt{KernelSearch}$(K, \phi, \bbM, \mD)$}
\label{alg_full}
\If{$\phi = -1$}{
        \Return $K$;
}
Compute check matrix $H_K^{(\phi)}$;\\
$\bbS_\phi \gets \emptyset$;\\
\For{\textbf{each} $v \in \bbM_\phi$}{
        $s \gets v \cdot (H_K^{(\phi +1)})^{\top}$;\tcp{Compute syndrome}
        \If{$s \in \bbS_\phi$}{
        \tcp{We already processed this coset}
                \textbf{continue};
        }
        \Else{
                $\bbS_\phi \gets \bbS_\phi \cup \set{s}$
        }
        \tcp{Compute partial distance and weight distribution of $\mC_{K}^{(\phi+1)} \oplus v$
} 
        $(d, A_{K}^{(\phi+1)}(v)) \gets d_H(v, \mC_{K}^{(\phi+1)})$;

        \If{$d = \mD_\phi$}{
                \If{ $A_{K}^{(\phi+1)}(v) \in \bW_\phi$ }{
                        \tcp{We already processed coset with same $A_{K}^{(\phi+1)}(v)$}
                        \textbf{continue};\\
                }
                \Else{
                     $\bW_\phi \gets \bW_\phi \cup \set{A_{K}^{(\phi+1)}(v)}$
                }
                $K[\phi] \gets v$;\\
                $\widehat K \gets $ \texttt{KernelSearch}$(K, \phi-1, \bbM, \mD)$;\\
                \If{$\widehat K \ne \mathbf 0^{l \times l}$}{
                        $\bK \gets \bK \cup \set{\widehat K}$
                }
        }
}
\Return{$\mathbf 0^{l \times l}$}
\end{algorithm}

%% file: maximization.tex
\section{Maximizing rate of polarization}
\label{sMaximization}

\begin{table*}[htbp]
\caption{Polarization properties of the kernels with the best rate of polarization}
\label{tableMax}
\footnotesize
\begin{tabular}{
|@{\hspace{1mm}}>{\centering}p{2.3mm}@{\hspace{1mm}}
|@{\hspace{1mm}}>{\centering}p{10mm}@{\hspace{1mm}}
|@{\hspace{1mm}}>{\centering}p{9mm}@{\hspace{1mm}}
||@{\hspace{1mm}}>{\centering}p{10mm}@{\hspace{1mm}}
|@{\hspace{1mm}}>{\centering}p{9mm}@{\hspace{1mm}}
|l
|c
|c|}
\hline
\multirow{3}{*}{$l$} & \multirow{3}{*}{\cite{korada2010polar}, $E$} & \multirow{3}{*}{\cite{presman2015binary}, $E$} & \multicolumn{5}{c|}{Proposed}                                                                                                                                      \\ \cline{4-8} 
                     &                                              &                                                & \multirow{2}{*}{$E$} & \multirow{2}{*}{$\mu$} & \multicolumn{1}{c|}{\multirow{2}{*}{Partial Distances}} & \multicolumn{2}{c|}{Processing complexity}               \\ \cline{7-8} 
                     &                                              &                                                &                      &                        & \multicolumn{1}{c|}{}                                   & \multicolumn{1}{c|}{RTPA} & \multicolumn{1}{c|}{Viterbi} \\ \hline
17                   & 0.49175                      &                             & \textbf{0.49361}       & 3.57316       & 1, 1, 2, 2, 2, 3, 4, 4, 4, 5, 6, 7, 8, 8, 8, 8, 16                                        & 1250                       & 6938            \\ \hline
18                   & 0.48968                      & 0.49521                     & \textbf{0.50052}       & 3.52842       & 1, 2, 2, 2, 2, 2, 4,   4, 4, 6, 6, 6, 6, 8, 8, 10, 10, 12                                 & 2946                       & 17414           \\ \hline
19                   & 0.48742                      & 0.49045                     & \textbf{0.50054}       & 3.44434       & 1, 2, 2, 2, 2, 2, 4,   4, 4, 4, 6, 6, 6, 8, 8, 8, 10, 10, 16                              & 5048                       & 17754           \\ \hline
20                   & 0.49659                      &                             & \textbf{0.50617}       & 3.43827       & 1, 2, 2, 2, 2, 2, 4,   4, 4, 4, 6, 6, 8, 8, 8, 8, 8, 8, 12, 16                            & 4894                       & 22578           \\ \hline
21                   & 0.48705                      & 0.49604                     & \textbf{0.50868}       & 3.37385       & 1, 2, 2, 2, 2, 2, 4,   4, 4, 4, 6, 6, 6, 6, 8, 8, 10, 10, 10, 14, 14                      & 10978                      & 39630           \\ \hline
22                   & 0.49445                      & 0.50118                     & \textbf{0.51181}       & 3.3530       & 1, 2, 2, 2, 2, 2, 4,   4, 4, 4, 6, 6, 6, 6, 8, 8, 8, 10, 10, 10, 12, 20                   & 18120                      & 70430           \\ \hline
23                   & 0.50071                      & 0.50705                     & \textbf{0.51213}      & 3.37214       & 1, 2, 2, 2, 2, 2, 4,   4, 4, 4, 6, 6, 6, 6, 6, 8, 8, 10, 10, 10, 12, 14, 16               & 17786                      & 67946           \\ \hline
24                   & 0.50445                      & 0.51577                     & 0.51577       & 3.3113       & 1, 2, 2, 2, 2, 2, 4,   4, 4, 4, 4, 4, 8, 8, 8, 8, 8, 8, 8, 12, 12, 12, 16, 16             & 2828         

           & 29782           \\ \hline
25                   & 0.50040                      & 0.50608                     & \textbf{0.51683}       & 3.28057       & 1, 2, 2, 2, 2, 2, 4,   4, 4, 4, 4, 6, 6, 6, 8, 8, 8, 8, 8, 10, 12, 12, 12, 16, 18         & 40566                      & 164943          \\ \hline
26                   & 0.50470                      &                             & \textbf{0.51921}       & 3.25551       & 1, 2, 2, 2, 2, 2, 4,   4, 4, 4, 4, 6, 6, 6, 6, 8, 8, 8, 10, 10, 10, 12, 12, 12, 14, 24    & 54848                      & 191819          \\ \hline
27                   & 0.50836                      &                             & \textbf{0.51935}       & 3.27814       & 1, 2, 2, 2, 2, 2, 4,   4, 4, 4, 4, 6, 6, 6, 6, 8, 8, 8, 8, 10, 10, 10, 12, 12, 14, 14, 24 & 93764                      & 327428          \\ \hline
\end{tabular}
\end{table*}

In this section we describe a method of polarization kernel search, which in many cases improves the lower bound on the maximal rate of polarization $E_l$ achievable by $l \times l$ polarization kernel, i.e.  $E_l  = \max_{K \in \bF_2^{l \times l}} E(K)$. Lower and upper bounds on $E_l$ for some $l$ were derived in \cite{korada2010polar}, \cite{presman2015binary} and 
\cite{lin2015linear} together with explicit kernels achieving lower bounds.
\subsection{Search for good partial distance profiles}
\label{ssBounds}
According to \cite[Corollary 16]{korada2010polar}, maximization of $E_l$ can be done by considering only kernels with nondecreasing partial distances, i.e. $\mD_\phi \leq \mD_{\phi+1}$ for $\phi \in [l-1]$.

\begin{proposition}[Minimum Distance and Partial Distance \cite{korada2010polar}]
\label{pMinPartD}
Let $\mC_0$ be a $(n,k, d_0)$ binary linear code. Let $g$ be an length-$n$ vector  with $d_H(g,\mC_0) = d_2$. Let $\mC_1$ be the $(n,k+1,d_1)$ linear code obtained by adding the vector $g$ to $\mC_0$, i.e., $\mC_1 = \langle g, \mC_0 \rangle$. Then $d_ 2 = \min(d_0,d_1)$.
\end{proposition}

Consider $l \times l$ kernel $K$ with PDP $\mD$.
Proposition \ref{pMinPartD} implies that the minimum distance $d_K^{(\phi)}$ of kernel codes $\mathcal C_K^{(\phi)}$, $\phi \in [l]$ is given by $\min_{\phi \leq i < l} \mD_i$. Let $d[n,k]$ be a best known minimum distance of $(n, k)$ linear code \cite{Grassl:codetables}. Thus, we can apply bound $d[l,l-\phi]$ on kernel codes $\mathcal C_K^{(\phi)}$, which results in the restriction  on partial distance $\mD_\phi \leq d[l,l-\phi]$.

\begin{lemma}(\cite[Lemma 4]{lin2015linear})
\label{lLin4}
Let $\mD_0, \mD_1, \dots, \mD_{l-1}$ be a nondecreasing partial distance profile of a kernel of size $l$. If $\mD_1 = 2$, then $\mD_i$ is even for all $i \geq 1$.
\end{lemma}

\begin{lemma}(\cite[Lemma 5]{lin2015linear})
\label{lLin5}
Let $\mD_0, \mD_1, \dots, \mD_{l-1}$ be a nondecreasing partial distance profile of a kernel $K$ of size $l$. Then, for $0 \leq i < l$, we have
$
\sum_{i' = i}^{l}2^{l-i'}\mD_{i'} \leq 2^{l-i} l.
$
\end{lemma}

To improve $E_l$, we propose the approach similar to those described in \cite{lin2015linear}. Namely, we generate nondecreasing PDPs $\mD$ with $\mD_{\phi}  \leq d[l,l-\phi]$ satisfying Lemmas \ref{lLin4}, \ref{lLin5}, and which are LP-valid \cite{presman2015binary}. We also require that $\frac{1}{l} \sum_{i=0}^{l-1} \log_l \mD_i$ should be larger than existing lower bounds on $E_l$. Then, we run Alg. \ref{alg_full} for all valid $\mD$. Due to the tremendous search space, for some PDP $\mD$ the algorithm may not produce a kernel for a long time. In this case, it should be terminated. Nevertheless, during the computer search we observed that if a kernel with given PDP\ exists, Alg. \ref{alg_full} often returns it quickly (in a few seconds). This is the advantage of depth-first search, which allows us to perform maximization of $E_l$.

\subsection{Results}
Table \ref{tableMax} presents the parameters of the kernels obtained with the proposed approach. The proposed kernels can be found in
Appendix, Fig \ref{fBestKernels}.
It can be seen that the proposed kernels have better rate of polarization compared with the ones proposed in \cite{korada2010polar} and \cite{presman2015binary}, except for $l = 24$. 

We also report the processing complexity (measured as a number of addition and comparison operations) of the proposed kernels for Viterbi-based processing algorithm \cite{griesser2002aposteriori} and the recently proposed recursive trellis processing algorithm (RTPA) \cite{trifonov2021recursive}, \cite{trifonov2019trellis}. It can be seen that the complexity of RTPA is significantly lower compared with Viterbi processing. Note that we  did not optimize the processing complexity of the
proposed kernels, like it was done in \cite{moskovskaya2020design} to reduce the complexity of
Viterbi processing algorithm of BCH kernels.  

We did not try to minimize the scaling exponent of the kernels of considered sizes. We conjecture that it can be further decreased.

%% file: complexity.tex
\section{Kernels with low processing complexity}
\label{sComplexity}

\begin{table*}[htbp]
\caption{Polarization properties of kernels which admit low complexity processing}
\label{tableCompl}
\centering
\footnotesize
\begin{tabular}{
|@{\hspace{1mm}}>{\centering}p{2.3mm}@{\hspace{1mm}}
|@{\hspace{1mm}}>{\centering}p{10mm}@{\hspace{1mm}}
|@{\hspace{1mm}}>{\centering}p{9mm}@{\hspace{1mm}}
|l|c|c|}
\hline
\multirow{2}{*}{$l$} & \multirow{2}{*}{$E$} & \multirow{2}{*}{$\mu$} & \multicolumn{1}{c|}{\multirow{2}{*}{Partial distances}}                                                     & \multicolumn{2}{c|}{Processing complexity} \\ \cline{5-6} 
                     &                      &                        & \multicolumn{1}{c|}{}                                                                                       & RTPA         & Viterbi        \\ \hline
18                   & 0.49521              & 3.64809                & 1, 2, 2, 4, 2, 2, 2,   4, 4, 6, 4, 6, 8, 8, 8, 8, 8, 16                                                     & 1000                      & 6507           \\ \hline
20                   & 0.49943              & 3.64931                & 1, 2, 2, 4, 2, 4, 2,   2, 4, 4, 6, 8, 8, 8, 4, 8, 12, 8, 8, 16                                              & 478                       & 5632           \\ \hline
24                   & 0.50291              & 3.61903                & 1, 2, 2, 4, 2, 4, 2,   4, 6, 2, 4, 4, 8, 8, 12, 4, 4, 8, 8, 12, 12, 8, 16, 16                               & 365                       & 6320           \\ \hline
27                   & 0.49720              & 3.76596                & 1, 2, 2, 4, 2, 2, 4,   4, 6, 2, 4, 4, 6, 6, 8, 8, 10, 12, 4, 4, 8, 8, 12, 12, 8, 16, 16                     & 998                       & 14905          \\ \hline
32                   & 0.52194              & 3.42111                & 1, 2, 2, 4, 2, 4, 2,   4, 6, 8, 2, 4, 6, 8, 4, 6, 8, 12, 4, 8, 12, 16, 4, 8, 12, 16, 8, 16, 8, 16,   16, 32 & 526                       & 13608          \\ \hline
\end{tabular}
\end{table*}
Although the polarization kernels from Table \ref{tableMax} have excellent polarization properties, their practical usage is limited due to the tremendous processing complexity. We propose to consider polarization kernels with degraded polarization properties and non-monotonic partial distances. This is motivated by fact that the trellis state complexity of an $(n, k, d)$ linear code is lower bounded. That is, by degrading the rate of polarization and permuting the partial distances we reduce the minimal distance of the corresponding kernel codes, thus reducing the lower bound on trellis state complexity. For example, such kernels of size 16 and 32 were proposed in \cite{trofimiuk2019construction32} and \cite{trofimiuk2021window}.

We propose to use the algorithm given in Section \ref{ssBounds} to obtain several PDPs corresponding to a reduced rate of polarization. Then we permute some PDP entries and try to construct the corresponding kernels using Alg. \ref{alg_full}.

The results of complexity minimization are presented in Table \ref{tableCompl}. The proposed kernels can be found in 
Appendix, Fig \ref{fSimpleKernels}.
It can be seen that by degrading polarization properties  we can  significantly reduce the processing complexity. For instance, $E(K_{27}^\ast) = 0.51935$ and its RTPA complexity is $93764$ operations, whereas $E(K_{27}) = 0.49720$ and its processing complexity is $998$ operations, which is 94 times lower.  

%% file: numeric.tex
\section{Numeric results}
Several mixed-kernel (MK) polar subcodes with obtained kernels were constructed. Their performance was investigated for the case of AWGN\ channel with BPSK\ modulation. The sets of frozen symbols were obtained by the method proposed in \cite{trifonov2019construction}. For  MK polar subcodes we used different order of kernels and present the results for the best one.

%

\begin{figure}[htbp]
\resizebox{!}{0.325\textwidth}{\input{plots/768_384_l8.tex}}
\caption{Performance of $(768, 384)$ polar codes}
\label{f768}
\end{figure}
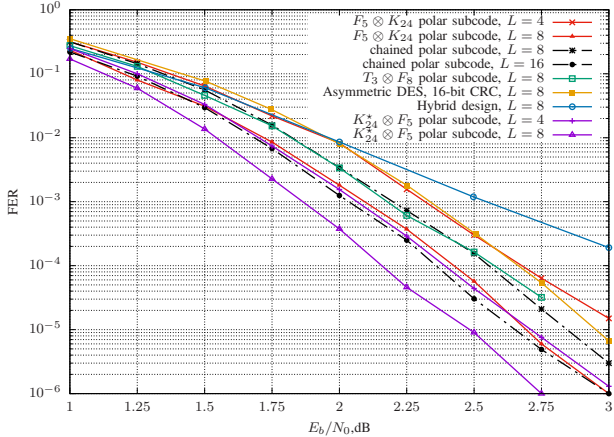

Fig. \ref{f768} illustrates the  performance of different $(768, 384)$ polar codes decoded by successive cancellation list (SCL) algorithm \cite{tal2015list} with list size $L$. We report the results for randomized MK polar subcodes \cite{trifonov2017randomized}, \cite{trifonov2019construction} with the proposed $K_{24}$ and $K_{24}^\ast$ kernels, $2^t \times 2^t$ Arikan matrix $
F_t = \left(
\arraycolsep=1.15pt\def\arraystretch{0.5}
\begin{array}{cc}
1&0\\
1&1
\end{array} \right)^{\otimes t}.
$
 and $3\times 3$ kernel $T_3$ introduced in \cite{bioglio2020multikernel}. We also include the results for descending (DES) asymmetric construction with 16-bit CRC \cite{cavatassi2019asymmetric} and improved hybrid design \cite{bioglio2019improved} of MK polar codes together with the performance of randomized chained polar subcodes \cite{trifonov2018randomized}. It can be seen that polar codes with $F_5 \otimes K_{24}$ polarizing transform under SCL decoding with $L = 8$ outperforms chained and $T_3 \otimes F_8$ polar subcodes, asymmetric DES and hybrid construction under SCL with the same list size. Moreover, it provides almost the same performance as chained polar subcode decoded with $L = 16$. Observe that despite of having degraded polarization properties (compared with $K_{24}^\ast$ kernel), kernel $K_{24}$ still provides a noticeable performance gain compared with other code constructions.


It can be seen that due to excellent polarization properties of $K_{24}^\ast$, MK polar code with  $K_{24}^{\ast} \otimes  F_5$ under SCL with $L = 4$ provides the same performance as $F_5 \otimes K_{24}$ with $L = 8$. 

\begin{figure}[htbp]
\centering
\resizebox{!}{0.325\textwidth}{\input{plots/1944_972_L8.tex}}
\caption{Performance of $(1944, 972)$ polar codes}
\label{f1944}
\end{figure}
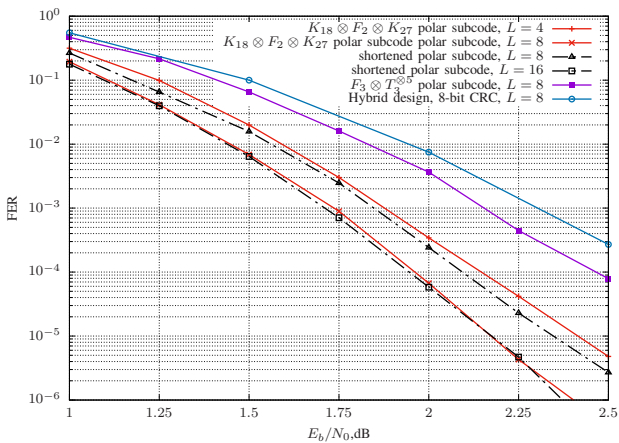

Fig. \ref{f1944} presents the performance of various $(1944, 972)$ polar codes. It can be seen that polar subcode with $K_{18} \otimes F_{2} \otimes K_{27}$ polarizing transform under SCL decoding with $L = 4$ provides significant performance gain compared to MK polar subcode with $F_3 \otimes T_3^{\otimes 5}$ polarizing transform and hybrid design wit 8-bit CRC \cite{bioglio2020multikernel}. Moreover,  $K_{18} \otimes F_{2} \otimes K_{27}$ polar subcode with $L = 8$ has the same performance as shortened polar subcode decoded with $L = 16$. We should also point out that despite of $E(K_{18}) <\ 0.5$ and $E(K_{27}) <\ 0.5$, the resulting code achieves significant performance gain compared with shortened polar code with $F_{11}$ having $E(F_{11}) = 0.5$. In addition, we conjecture that polarization properties of kernels with different size can not be properly compared. That is, kernels $K$ with $l \ne 2^t$ and even $E(K) <\ 0.5$ should be studied and can be considered in mixed kernel polar codes.

More results on performance and complexity comparison of polar codes with the proposed kernels can be found in \cite{trifonov2021recursive}.


%% file: plots/768_384_l8.tex
\begingroup
  \inputencoding{cp1251}%
\footnotesize
  \makeatletter
  \providecommand\color[2][]{%
    \GenericError{(gnuplot) \space\space\space\@spaces}{%
      Package color not loaded in conjunction with
      terminal option `colourtext'%
    }{See the gnuplot documentation for explanation.%
    }{Either use 'blacktext' in gnuplot or load the package
      color.sty in LaTeX.}%
    \renewcommand\color[2][]{}%
  }%
  \providecommand\includegraphics[2][]{%
    \GenericError{(gnuplot) \space\space\space\@spaces}{%
      Package graphicx or graphics not loaded%
    }{See the gnuplot documentation for explanation.%
    }{The gnuplot epslatex terminal needs graphicx.sty or graphics.sty.}%
    \renewcommand\includegraphics[2][]{}%
  }%
  \providecommand\rotatebox[2]{#2}%
  \@ifundefined{ifGPcolor}{%
    \newif\ifGPcolor
    \GPcolortrue
  }{}%
  \@ifundefined{ifGPblacktext}{%
    \newif\ifGPblacktext
    \GPblacktexttrue
  }{}%
  \let\gplgaddtomacro\g@addto@macro
  \gdef\gplbacktext{}%
  \gdef\gplfronttext{}%
  \makeatother
  \ifGPblacktext
    \def\colorrgb#1{}%
    \def\colorgray#1{}%
  \else
    \ifGPcolor
      \def\colorrgb#1{\color[rgb]{#1}}%
      \def\colorgray#1{\color[gray]{#1}}%
      \expandafter\def\csname LTw\endcsname{\color{white}}%
      \expandafter\def\csname LTb\endcsname{\color{black}}%
      \expandafter\def\csname LTa\endcsname{\color{black}}%
      \expandafter\def\csname LT0\endcsname{\color[rgb]{1,0,0}}%
      \expandafter\def\csname LT1\endcsname{\color[rgb]{0,1,0}}%
      \expandafter\def\csname LT2\endcsname{\color[rgb]{0,0,1}}%
      \expandafter\def\csname LT3\endcsname{\color[rgb]{1,0,1}}%
      \expandafter\def\csname LT4\endcsname{\color[rgb]{0,1,1}}%
      \expandafter\def\csname LT5\endcsname{\color[rgb]{1,1,0}}%
      \expandafter\def\csname LT6\endcsname{\color[rgb]{0,0,0}}%
      \expandafter\def\csname LT7\endcsname{\color[rgb]{1,0.3,0}}%
      \expandafter\def\csname LT8\endcsname{\color[rgb]{0.5,0.5,0.5}}%
    \else
      \def\colorrgb#1{\color{black}}%
      \def\colorgray#1{\color[gray]{#1}}%
      \expandafter\def\csname LTw\endcsname{\color{white}}%
      \expandafter\def\csname LTb\endcsname{\color{black}}%
      \expandafter\def\csname LTa\endcsname{\color{black}}%
      \expandafter\def\csname LT0\endcsname{\color{black}}%
      \expandafter\def\csname LT1\endcsname{\color{black}}%
      \expandafter\def\csname LT2\endcsname{\color{black}}%
      \expandafter\def\csname LT3\endcsname{\color{black}}%
      \expandafter\def\csname LT4\endcsname{\color{black}}%
      \expandafter\def\csname LT5\endcsname{\color{black}}%
      \expandafter\def\csname LT6\endcsname{\color{black}}%
      \expandafter\def\csname LT7\endcsname{\color{black}}%
      \expandafter\def\csname LT8\endcsname{\color{black}}%
    \fi
  \fi
    \setlength{\unitlength}{0.0500bp}%
    \ifx\gptboxheight\undefined%
      \newlength{\gptboxheight}%
      \newlength{\gptboxwidth}%
      \newsavebox{\gptboxtext}%
    \fi%
    \setlength{\fboxrule}{0.5pt}%
    \setlength{\fboxsep}{1pt}%
\begin{picture}(7200.00,5040.00)%
    \gplgaddtomacro\gplbacktext{%
      \csname LTb\endcsname
      \put(688,512){\makebox(0,0)[r]{\strut{}$10^{-6}$}}%
      \csname LTb\endcsname
      \put(688,1240){\makebox(0,0)[r]{\strut{}$10^{-5}$}}%
      \csname LTb\endcsname
      \put(688,1968){\makebox(0,0)[r]{\strut{}$10^{-4}$}}%
      \csname LTb\endcsname
      \put(688,2696){\makebox(0,0)[r]{\strut{}$10^{-3}$}}%
      \csname LTb\endcsname
      \put(688,3423){\makebox(0,0)[r]{\strut{}$10^{-2}$}}%
      \csname LTb\endcsname
      \put(688,4151){\makebox(0,0)[r]{\strut{}$10^{-1}$}}%
      \csname LTb\endcsname
      \put(688,4879){\makebox(0,0)[r]{\strut{}$10^{0}$}}%
      \csname LTb\endcsname
      \put(784,352){\makebox(0,0){\strut{}$1$}}%
      \csname LTb\endcsname
      \put(1550,352){\makebox(0,0){\strut{}$1.25$}}%
      \csname LTb\endcsname
      \put(2316,352){\makebox(0,0){\strut{}$1.5$}}%
      \csname LTb\endcsname
      \put(3082,352){\makebox(0,0){\strut{}$1.75$}}%
      \csname LTb\endcsname
      \put(3848,352){\makebox(0,0){\strut{}$2$}}%
      \csname LTb\endcsname
      \put(4613,352){\makebox(0,0){\strut{}$2.25$}}%
      \csname LTb\endcsname
      \put(5379,352){\makebox(0,0){\strut{}$2.5$}}%
      \csname LTb\endcsname
      \put(6145,352){\makebox(0,0){\strut{}$2.75$}}%
      \csname LTb\endcsname
      \put(6911,352){\makebox(0,0){\strut{}$3$}}%
    }%
    \gplgaddtomacro\gplfronttext{%
      \csname LTb\endcsname
      \put(152,2695){\rotatebox{-270}{\makebox(0,0){\strut{}FER}}}%
      \put(3847,112){\makebox(0,0){\strut{}$E_b/N_0$,dB}}%
      \csname LTb\endcsname
      \put(6176,4736){\makebox(0,0)[r]{\strut{}$F_{5}\otimes K_{24}$ polar subcode, $L = 4$}}%
      \csname LTb\endcsname
      \put(6176,4576){\makebox(0,0)[r]{\strut{}$F_{5}\otimes K_{24}$ polar subcode, $L = 8$}}%
      \csname LTb\endcsname
      \put(6176,4416){\makebox(0,0)[r]{\strut{}chained polar subcode, $L = 8$}}%
      \csname LTb\endcsname
      \put(6176,4256){\makebox(0,0)[r]{\strut{}chained polar subcode, $L = 16$}}%
      \csname LTb\endcsname
      \put(6176,4096){\makebox(0,0)[r]{\strut{}$T_{3} \otimes F_{8}$ polar subcode, $L = 8$}}%
      \csname LTb\endcsname
      \put(6176,3936){\makebox(0,0)[r]{\strut{}Asymmetric DES, 16-bit CRC, $L = 8$}}%
      \csname LTb\endcsname
      \put(6176,3776){\makebox(0,0)[r]{\strut{}Hybrid design, $L = 8$}}%
      \csname LTb\endcsname
      \put(6176,3616){\makebox(0,0)[r]{\strut{}$K_{24}^{\star} \otimes F_{5}$ polar subcode, $L = 4$}}%
      \csname LTb\endcsname
      \put(6176,3456){\makebox(0,0)[r]{\strut{}$K_{24}^{\star} \otimes F_{5}$ polar subcode, $L = 8$}}%
    }%
    \gplbacktext
    \put(0,0){\includegraphics[width={360.00bp},height={252.00bp}]{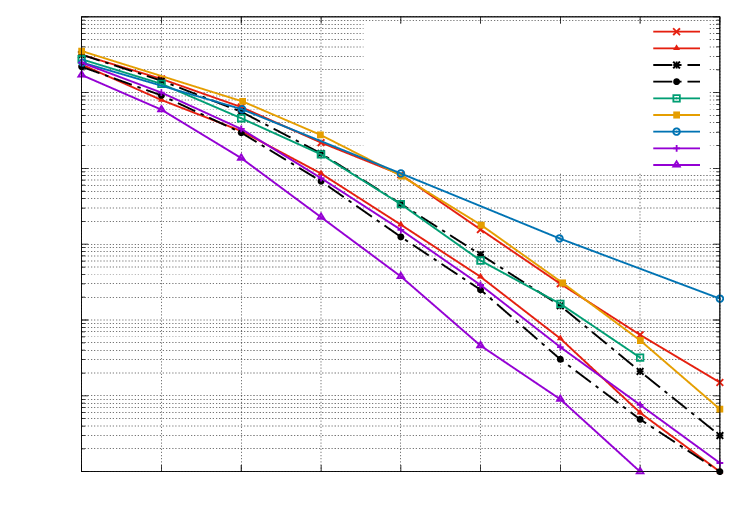}}%
    \gplfronttext
  \end{picture}%
\endgroup

%% file: plots/1944_972_L8.tex
\begingroup
  \inputencoding{cp1251}%
\footnotesize
  \makeatletter
  \providecommand\color[2][]{%
    \GenericError{(gnuplot) \space\space\space\@spaces}{%
      Package color not loaded in conjunction with
      terminal option `colourtext'%
    }{See the gnuplot documentation for explanation.%
    }{Either use 'blacktext' in gnuplot or load the package
      color.sty in LaTeX.}%
    \renewcommand\color[2][]{}%
  }%
  \providecommand\includegraphics[2][]{%
    \GenericError{(gnuplot) \space\space\space\@spaces}{%
      Package graphicx or graphics not loaded%
    }{See the gnuplot documentation for explanation.%
    }{The gnuplot epslatex terminal needs graphicx.sty or graphics.sty.}%
    \renewcommand\includegraphics[2][]{}%
  }%
  \providecommand\rotatebox[2]{#2}%
  \@ifundefined{ifGPcolor}{%
    \newif\ifGPcolor
    \GPcolortrue
  }{}%
  \@ifundefined{ifGPblacktext}{%
    \newif\ifGPblacktext
    \GPblacktexttrue
  }{}%
  \let\gplgaddtomacro\g@addto@macro
  \gdef\gplbacktext{}%
  \gdef\gplfronttext{}%
  \makeatother
  \ifGPblacktext
    \def\colorrgb#1{}%
    \def\colorgray#1{}%
  \else
    \ifGPcolor
      \def\colorrgb#1{\color[rgb]{#1}}%
      \def\colorgray#1{\color[gray]{#1}}%
      \expandafter\def\csname LTw\endcsname{\color{white}}%
      \expandafter\def\csname LTb\endcsname{\color{black}}%
      \expandafter\def\csname LTa\endcsname{\color{black}}%
      \expandafter\def\csname LT0\endcsname{\color[rgb]{1,0,0}}%
      \expandafter\def\csname LT1\endcsname{\color[rgb]{0,1,0}}%
      \expandafter\def\csname LT2\endcsname{\color[rgb]{0,0,1}}%
      \expandafter\def\csname LT3\endcsname{\color[rgb]{1,0,1}}%
      \expandafter\def\csname LT4\endcsname{\color[rgb]{0,1,1}}%
      \expandafter\def\csname LT5\endcsname{\color[rgb]{1,1,0}}%
      \expandafter\def\csname LT6\endcsname{\color[rgb]{0,0,0}}%
      \expandafter\def\csname LT7\endcsname{\color[rgb]{1,0.3,0}}%
      \expandafter\def\csname LT8\endcsname{\color[rgb]{0.5,0.5,0.5}}%
    \else
      \def\colorrgb#1{\color{black}}%
      \def\colorgray#1{\color[gray]{#1}}%
      \expandafter\def\csname LTw\endcsname{\color{white}}%
      \expandafter\def\csname LTb\endcsname{\color{black}}%
      \expandafter\def\csname LTa\endcsname{\color{black}}%
      \expandafter\def\csname LT0\endcsname{\color{black}}%
      \expandafter\def\csname LT1\endcsname{\color{black}}%
      \expandafter\def\csname LT2\endcsname{\color{black}}%
      \expandafter\def\csname LT3\endcsname{\color{black}}%
      \expandafter\def\csname LT4\endcsname{\color{black}}%
      \expandafter\def\csname LT5\endcsname{\color{black}}%
      \expandafter\def\csname LT6\endcsname{\color{black}}%
      \expandafter\def\csname LT7\endcsname{\color{black}}%
      \expandafter\def\csname LT8\endcsname{\color{black}}%
    \fi
  \fi
    \setlength{\unitlength}{0.0500bp}%
    \ifx\gptboxheight\undefined%
      \newlength{\gptboxheight}%
      \newlength{\gptboxwidth}%
      \newsavebox{\gptboxtext}%
    \fi%
    \setlength{\fboxrule}{0.5pt}%
    \setlength{\fboxsep}{1pt}%
\begin{picture}(7200.00,5040.00)%
    \gplgaddtomacro\gplbacktext{%
      \csname LTb\endcsname
      \put(688,512){\makebox(0,0)[r]{\strut{}$10^{-6}$}}%
      \csname LTb\endcsname
      \put(688,1240){\makebox(0,0)[r]{\strut{}$10^{-5}$}}%
      \csname LTb\endcsname
      \put(688,1968){\makebox(0,0)[r]{\strut{}$10^{-4}$}}%
      \csname LTb\endcsname
      \put(688,2696){\makebox(0,0)[r]{\strut{}$10^{-3}$}}%
      \csname LTb\endcsname
      \put(688,3423){\makebox(0,0)[r]{\strut{}$10^{-2}$}}%
      \csname LTb\endcsname
      \put(688,4151){\makebox(0,0)[r]{\strut{}$10^{-1}$}}%
      \csname LTb\endcsname
      \put(688,4879){\makebox(0,0)[r]{\strut{}$10^{0}$}}%
      \csname LTb\endcsname
      \put(784,352){\makebox(0,0){\strut{}$1$}}%
      \csname LTb\endcsname
      \put(1805,352){\makebox(0,0){\strut{}$1.25$}}%
      \csname LTb\endcsname
      \put(2826,352){\makebox(0,0){\strut{}$1.5$}}%
      \csname LTb\endcsname
      \put(3848,352){\makebox(0,0){\strut{}$1.75$}}%
      \csname LTb\endcsname
      \put(4869,352){\makebox(0,0){\strut{}$2$}}%
      \csname LTb\endcsname
      \put(5890,352){\makebox(0,0){\strut{}$2.25$}}%
      \csname LTb\endcsname
      \put(6911,352){\makebox(0,0){\strut{}$2.5$}}%
    }%
    \gplgaddtomacro\gplfronttext{%
      \csname LTb\endcsname
      \put(152,2695){\rotatebox{-270}{\makebox(0,0){\strut{}FER}}}%
      \put(3847,112){\makebox(0,0){\strut{}$E_b/N_0$,dB}}%
      \csname LTb\endcsname
      \put(6176,4736){\makebox(0,0)[r]{\strut{}$K_{18} \otimes F_{2} \otimes K_{27}$ polar subcode, $L = 4$}}%
      \csname LTb\endcsname
      \put(6176,4576){\makebox(0,0)[r]{\strut{}$K_{18} \otimes F_{2} \otimes K_{27}$ polar subcode polar subcode, $L = 8$}}%
      \csname LTb\endcsname
      \put(6176,4416){\makebox(0,0)[r]{\strut{}shortened polar subcode, $L = 8$}}%
      \csname LTb\endcsname
      \put(6176,4256){\makebox(0,0)[r]{\strut{}shortened polar subcode, $L = 16$}}%
      \csname LTb\endcsname
      \put(6176,4096){\makebox(0,0)[r]{\strut{}$F_3\otimes T_3^{\otimes 5}$ polar subcode, $L = 8$}}%
      \csname LTb\endcsname
      \put(6176,3936){\makebox(0,0)[r]{\strut{}Hybrid design, 8-bit CRC, $L = 8$}}%
    }%
    \gplbacktext
    \put(0,0){\includegraphics[width={360.00bp},height={252.00bp}]{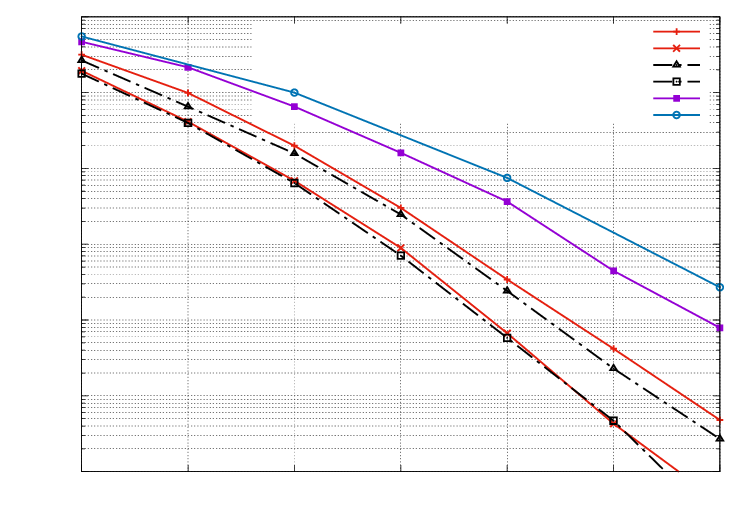}}%
    \gplfronttext
  \end{picture}%
\endgroup

%% file: conclusions.tex
\section{Conclusion and open problems}
In this paper a novel search method for large polarization kernels with given partial distance profile was proposed. This algorithm is based on depth-first search over sets of candidate rows. The proposed algorithm was used to improve lower bounds on the maximal rate of polarization of sizes $17 \leq l \leq 27$. 

We also demonstrated application of the proposed method for finding kernels which admit low complexity processing by the recursive trellis processing algorithm. Numerical results demonstrate the advantage of polar codes with the obtained kernels compared with shortened polar codes with Arikan kernel and polar codes with small kernels.

In the following we outline multiple open problems with large kernels:
\begin{itemize}
\item \textit{Tight bounds on partial distances.} There is a significant gap between lower and upper bounds on $E_l$ for large $l$. Moreover, we have almost no bounds for non-monotonic partial distances, except LP\ bounds.

\item The problem of \textit{scaling exponent minimization} remains unsolved. It is unknown whether the proposed method can be extended to explicitly solve it.
\item \textit{Processing complexity.} For some PDP the proposed algorithm can produce kernels which admit low complexity processing by RTPA. However, we do not know how to explicitly minimize a processing complexity for kernels with given PDP.

\section*{Acknowledgement} 
This work is partially supported by the Ministry of Science and Higher Education of Russian Federation, passport of goszadanie no. 2019-0898.

\end{itemize}

%% file: kernels.tex
\appendix
\begin{figure*}[htbp]
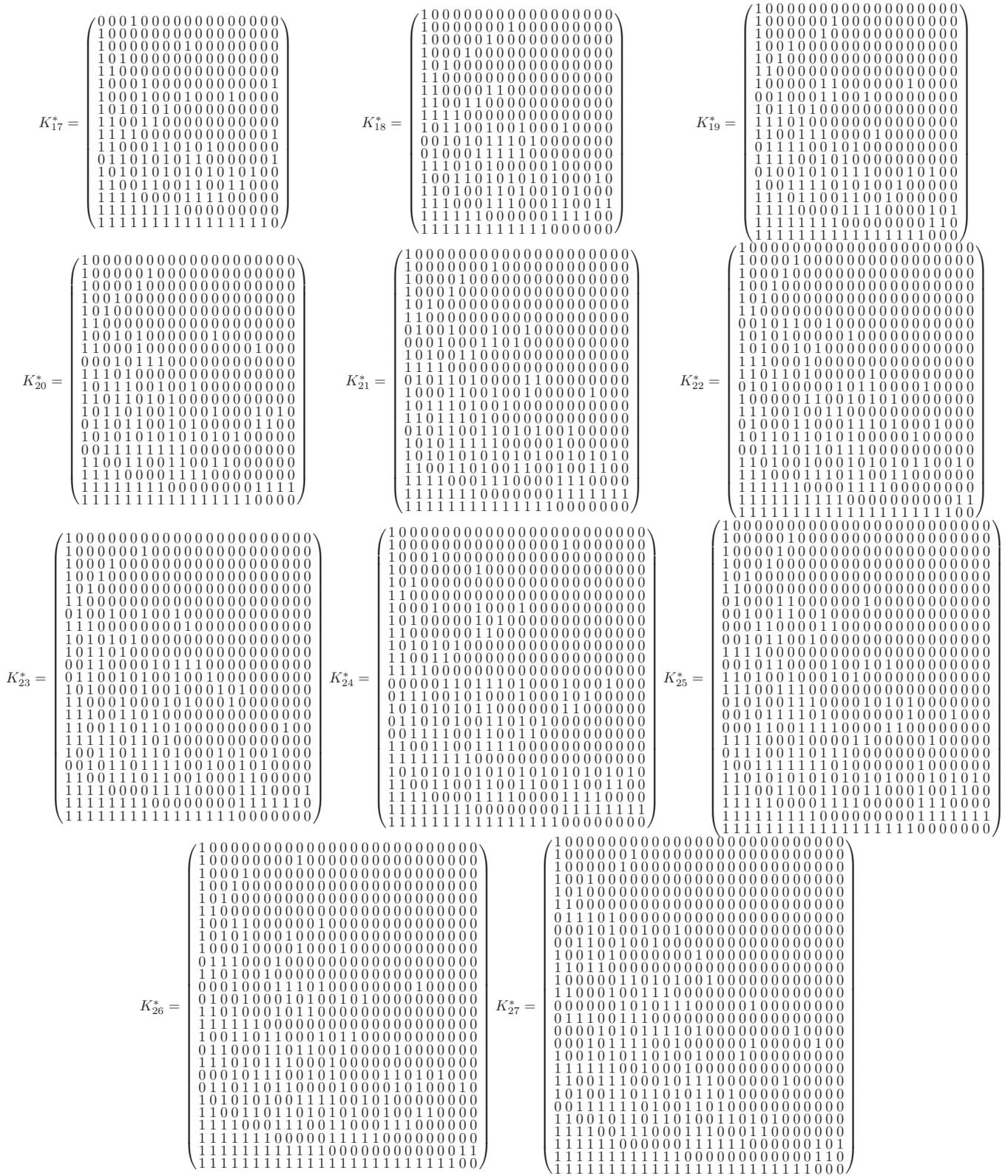

\centering
\scalebox{0.75}{
\parbox{1.7\textwidth}
{ 
$
\arraycolsep=1.2pt\def\arraystretch{0.6}
\begin{array}{ccc}
{
K_{17}^{\ast} = \left(\begin{array}{ccccccccccccccccc}
0&0&0&1&0&0&0&0&0&0&0&0&0&0&0&0&0\\
1&0&0&0&0&0&0&0&0&0&0&0&0&0&0&0&0\\
1&0&0&0&0&0&0&0&1&0&0&0&0&0&0&0&0\\
1&0&1&0&0&0&0&0&0&0&0&0&0&0&0&0&0\\
1&1&0&0&0&0&0&0&0&0&0&0&0&0&0&0&0\\
1&0&0&0&1&0&0&0&0&0&0&0&0&0&0&0&1\\
1&0&0&0&1&0&0&0&1&0&0&0&1&0&0&0&0\\
1&0&1&0&1&0&1&0&0&0&0&0&0&0&0&0&0\\
1&1&0&0&1&1&0&0&0&0&0&0&0&0&0&0&0\\
1&1&1&1&0&0&0&0&0&0&0&0&0&0&0&0&1\\
1&1&0&0&0&1&1&0&1&0&1&0&0&0&0&0&0\\
0&1&1&0&1&0&1&0&1&1&0&0&0&0&0&0&1\\
1&0&1&0&1&0&1&0&1&0&1&0&1&0&1&0&0\\
1&1&0&0&1&1&0&0&1&1&0&0&1&1&0&0&0\\
1&1&1&1&0&0&0&0&1&1&1&1&0&0&0&0&0\\
1&1&1&1&1&1&1&1&0&0&0&0&0&0&0&0&0\\
1&1&1&1&1&1&1&1&1&1&1&1&1&1&1&1&0\\
\end{array}\right)
}
&
{
K_{18}^{\ast} = \left(\begin{array}{cccccccccccccccccc}
1&0&0&0&0&0&0&0&0&0&0&0&0&0&0&0&0&0\\
1&0&0&0&0&0&0&0&1&0&0&0&0&0&0&0&0&0\\
1&0&0&0&0&0&1&0&0&0&0&0&0&0&0&0&0&0\\
1&0&0&0&1&0&0&0&0&0&0&0&0&0&0&0&0&0\\
1&0&1&0&0&0&0&0&0&0&0&0&0&0&0&0&0&0\\
1&1&0&0&0&0&0&0&0&0&0&0&0&0&0&0&0&0\\
1&1&0&0&0&0&1&1&0&0&0&0&0&0&0&0&0&0\\
1&1&0&0&1&1&0&0&0&0&0&0&0&0&0&0&0&0\\
1&1&1&1&0&0&0&0&0&0&0&0&0&0&0&0&0&0\\
1&0&1&1&0&0&1&0&0&1&0&0&0&1&0&0&0&0\\
0&0&1&0&1&0&1&1&1&0&1&0&0&0&0&0&0&0\\
0&1&0&0&0&1&1&1&1&1&0&0&0&0&0&0&0&0\\
1&1&1&0&1&0&1&0&0&0&0&0&1&0&0&0&0&0\\
1&0&0&1&1&0&1&0&1&0&1&0&1&0&0&0&1&0\\
1&1&0&1&0&0&1&1&0&1&0&0&1&0&1&0&0&0\\
1&1&1&0&0&0&1&1&1&0&0&0&1&1&0&0&1&1\\
1&1&1&1&1&1&0&0&0&0&0&0&1&1&1&1&0&0\\
1&1&1&1&1&1&1&1&1&1&1&1&0&0&0&0&0&0\\
\end{array}\right)
} 
& 
{
K_{19}^{\ast} = \left(\begin{array}{ccccccccccccccccccc}
1&0&0&0&0&0&0&0&0&0&0&0&0&0&0&0&0&0&0\\
1&0&0&0&0&0&0&1&0&0&0&0&0&0&0&0&0&0&0\\
1&0&0&0&0&0&1&0&0&0&0&0&0&0&0&0&0&0&0\\
1&0&0&1&0&0&0&0&0&0&0&0&0&0&0&0&0&0&0\\
1&0&1&0&0&0&0&0&0&0&0&0&0&0&0&0&0&0&0\\
1&1&0&0&0&0&0&0&0&0&0&0&0&0&0&0&0&0&0\\
1&0&0&0&0&0&1&1&0&0&0&0&0&0&1&0&0&0&0\\
0&0&1&0&0&0&1&1&0&0&1&0&0&0&0&0&0&0&0\\
1&0&1&1&0&1&0&0&0&0&0&0&0&0&0&0&0&0&0\\
1&1&1&0&1&0&0&0&0&0&0&0&0&0&0&0&0&0&0\\
1&1&0&0&1&1&1&0&0&0&0&1&0&0&0&0&0&0&0\\
0&1&1&1&1&0&0&1&0&1&0&0&0&0&0&0&0&0&0\\
1&1&1&1&0&0&1&0&1&0&0&0&0&0&0&0&0&0&0\\
0&1&0&0&1&0&1&0&1&1&1&0&0&0&1&0&1&0&0\\
1&0&0&1&1&1&1&0&1&0&1&0&0&1&0&0&0&0&0\\
1&1&1&0&1&1&0&0&1&1&0&0&1&0&0&0&0&0&0\\
1&1&1&1&0&0&0&0&1&1&1&1&0&0&0&0&1&0&1\\
1&1&1&1&1&1&1&1&0&0&0&0&0&0&0&0&1&1&0\\
1&1&1&1&1&1&1&1&1&1&1&1&1&1&1&1&0&0&0\\
\end{array}\right)
} 
\\
{
K_{20}^{\ast} = \left(\begin{array}{cccccccccccccccccccc}
1&0&0&0&0&0&0&0&0&0&0&0&0&0&0&0&0&0&0&0\\
1&0&0&0&0&0&1&0&0&0&0&0&0&0&0&0&0&0&0&0\\
1&0&0&0&0&1&0&0&0&0&0&0&0&0&0&0&0&0&0&0\\
1&0&0&1&0&0&0&0&0&0&0&0&0&0&0&0&0&0&0&0\\
1&0&1&0&0&0&0&0&0&0&0&0&0&0&0&0&0&0&0&0\\
1&1&0&0&0&0&0&0&0&0&0&0&0&0&0&0&0&0&0&0\\
1&0&0&1&0&1&0&0&0&0&0&0&1&0&0&0&0&0&0&0\\
1&1&0&0&0&1&0&0&0&0&0&0&0&0&0&0&1&0&0&0\\
0&0&0&1&0&1&1&1&0&0&0&0&0&0&0&0&0&0&0&0\\
1&1&1&0&1&0&0&0&0&0&0&0&0&0&0&0&0&0&0&0\\
1&0&1&1&1&0&0&1&0&0&1&0&0&0&0&0&0&0&0&0\\
1&1&0&1&1&0&1&0&1&0&0&0&0&0&0&0&0&0&0&0\\
1&0&1&1&0&1&0&0&1&0&0&0&1&0&0&0&1&0&1&0\\
0&1&1&0&1&1&0&0&1&0&1&0&0&0&0&0&1&1&0&0\\
1&0&1&0&1&0&1&0&1&0&1&0&1&0&1&0&0&0&0&0\\
0&0&1&1&1&1&1&1&1&1&0&0&0&0&0&0&0&0&0&0\\
1&1&0&0&1&1&0&0&1&1&0&0&1&1&0&0&0&0&0&0\\
1&1&1&1&0&0&0&0&1&1&1&1&0&0&0&0&0&0&0&0\\
1&1&1&1&1&1&1&1&0&0&0&0&0&0&0&0&1&1&1&1\\
1&1&1&1&1&1&1&1&1&1&1&1&1&1&1&1&0&0&0&0\\
\end{array}\right)
}
&
{
K_{21}^{\ast} = \left(\begin{array}{ccccccccccccccccccccc}
1&0&0&0&0&0&0&0&0&0&0&0&0&0&0&0&0&0&0&0&0\\
1&0&0&0&0&0&0&0&1&0&0&0&0&0&0&0&0&0&0&0&0\\
1&0&0&0&0&1&0&0&0&0&0&0&0&0&0&0&0&0&0&0&0\\
1&0&0&0&1&0&0&0&0&0&0&0&0&0&0&0&0&0&0&0&0\\
1&0&1&0&0&0&0&0&0&0&0&0&0&0&0&0&0&0&0&0&0\\
1&1&0&0&0&0&0&0&0&0&0&0&0&0&0&0&0&0&0&0&0\\
0&1&0&0&1&0&0&0&1&0&0&1&0&0&0&0&0&0&0&0&0\\
0&0&0&1&0&0&0&1&1&0&1&0&0&0&0&0&0&0&0&0&0\\
1&0&1&0&0&1&1&0&0&0&0&0&0&0&0&0&0&0&0&0&0\\
1&1&1&1&0&0&0&0&0&0&0&0&0&0&0&0&0&0&0&0&0\\
0&1&0&1&1&0&1&0&0&0&0&1&1&0&0&0&0&0&0&0&0\\
1&0&0&0&1&1&0&0&1&0&0&1&0&0&0&0&0&1&0&0&0\\
1&0&1&1&1&0&1&0&0&1&0&0&0&0&0&0&0&0&0&0&0\\
1&1&0&1&1&1&0&1&0&0&0&0&0&0&0&0&0&0&0&0&0\\
0&1&0&1&1&0&0&1&1&0&1&0&1&0&0&1&0&0&0&0&0\\
1&0&1&0&1&1&1&1&1&0&0&0&0&0&1&0&0&0&0&0&0\\
1&0&1&0&1&0&1&0&1&0&1&0&1&0&0&1&0&1&0&1&0\\
1&1&0&0&1&1&0&1&0&0&1&1&0&0&1&0&0&1&1&0&0\\
1&1&1&1&0&0&0&1&1&1&0&0&0&0&1&1&1&0&0&0&0\\
1&1&1&1&1&1&1&0&0&0&0&0&0&0&1&1&1&1&1&1&1\\
1&1&1&1&1&1&1&1&1&1&1&1&1&1&0&0&0&0&0&0&0\\
\end{array}\right)
} 
& 
{
K_{22}^{\ast} = \left(\begin{array}{cccccccccccccccccccccc}
1&0&0&0&0&0&0&0&0&0&0&0&0&0&0&0&0&0&0&0&0&0\\
1&0&0&0&0&1&0&0&0&0&0&0&0&0&0&0&0&0&0&0&0&0\\
1&0&0&0&1&0&0&0&0&0&0&0&0&0&0&0&0&0&0&0&0&0\\
1&0&0&1&0&0&0&0&0&0&0&0&0&0&0&0&0&0&0&0&0&0\\
1&0&1&0&0&0&0&0&0&0&0&0&0&0&0&0&0&0&0&0&0&0\\
1&1&0&0&0&0&0&0&0&0&0&0&0&0&0&0&0&0&0&0&0&0\\
0&0&1&0&1&1&0&0&1&0&0&0&0&0&0&0&0&0&0&0&0&0\\
1&0&1&0&1&0&0&0&0&0&1&0&0&0&0&0&0&0&0&0&0&0\\
1&0&1&0&0&1&0&1&0&0&0&0&0&0&0&0&0&0&0&0&0&0\\
1&1&1&0&0&0&1&0&0&0&0&0&0&0&0&0&0&0&0&0&0&0\\
1&1&0&1&1&0&1&0&0&0&0&0&1&0&0&0&0&0&0&0&0&0\\
0&1&0&1&0&0&0&0&0&1&0&1&1&0&0&0&0&1&0&0&0&0\\
1&0&0&0&0&0&1&1&0&0&1&0&1&0&1&0&0&0&0&0&0&0\\
1&1&1&0&0&1&0&0&1&1&0&0&0&0&0&0&0&0&0&0&0&0\\
0&1&0&0&0&1&1&0&0&0&1&1&1&0&1&0&0&0&1&0&0&0\\
1&0&1&1&0&1&1&0&1&0&1&0&0&0&0&0&1&0&0&0&0&0\\
0&0&1&1&1&0&1&1&0&1&1&1&0&0&0&0&0&0&0&0&0&0\\
1&1&0&1&0&0&1&0&0&0&1&0&1&0&1&0&1&1&0&0&1&0\\
1&1&1&0&0&0&1&1&1&0&1&1&0&0&1&1&0&0&0&0&0&0\\
1&1&1&1&1&1&0&0&0&0&1&1&1&1&0&0&0&0&0&0&0&0\\
1&1&1&1&1&1&1&1&1&1&0&0&0&0&0&0&0&0&0&0&1&1\\
1&1&1&1&1&1&1&1&1&1&1&1&1&1&1&1&1&1&1&1&0&0\\
\end{array}\right)
} 
\\
{
K_{23}^{\ast} = \left(\begin{array}{ccccccccccccccccccccccc}
1&0&0&0&0&0&0&0&0&0&0&0&0&0&0&0&0&0&0&0&0&0&0\\
1&0&0&0&0&0&0&1&0&0&0&0&0&0&0&0&0&0&0&0&0&0&0\\
1&0&0&0&1&0&0&0&0&0&0&0&0&0&0&0&0&0&0&0&0&0&0\\
1&0&0&1&0&0&0&0&0&0&0&0&0&0&0&0&0&0&0&0&0&0&0\\
1&0&1&0&0&0&0&0&0&0&0&0&0&0&0&0&0&0&0&0&0&0&0\\
1&1&0&0&0&0&0&0&0&0&0&0&0&0&0&0&0&0&0&0&0&0&0\\
0&1&0&0&1&0&0&1&0&0&1&0&0&0&0&0&0&0&0&0&0&0&0\\
1&1&1&0&0&0&0&0&0&0&0&1&0&0&0&0&0&0&0&0&0&0&0\\
1&0&1&0&1&0&1&0&0&0&0&0&0&0&0&0&0&0&0&0&0&0&0\\
1&0&1&1&0&1&0&0&0&0&0&0&0&0&0&0&0&0&0&0&0&0&0\\
0&0&1&1&0&0&0&0&1&0&1&1&1&0&0&0&0&0&0&0&0&0&0\\
0&1&1&0&0&1&0&1&0&0&1&0&0&1&0&0&0&0&0&0&0&0&0\\
1&0&1&0&0&0&0&1&0&0&1&0&0&0&1&0&1&0&0&0&0&0&0\\
1&1&0&0&0&1&0&0&0&1&0&1&0&0&0&1&0&0&0&0&0&0&0\\
1&1&1&0&0&1&1&0&1&0&0&0&0&0&0&0&0&0&0&0&0&0&0\\
1&1&0&0&1&1&0&1&1&0&1&0&0&0&0&0&0&0&0&0&1&0&0\\
1&1&1&1&1&0&1&1&0&1&0&0&0&0&0&0&0&0&0&0&0&0&0\\
1&0&0&1&1&0&1&1&1&0&1&0&0&0&1&0&1&0&0&1&0&0&0\\
0&0&1&0&1&1&0&1&1&1&1&0&0&1&0&0&1&0&1&0&0&0&0\\
1&1&0&0&1&1&1&0&1&1&0&0&1&0&0&0&1&1&0&0&0&0&0\\
1&1&1&1&0&0&0&0&1&1&1&1&0&0&0&0&1&1&1&0&0&0&1\\
1&1&1&1&1&1&1&1&0&0&0&0&0&0&0&0&1&1&1&1&1&1&0\\
1&1&1&1&1&1&1&1&1&1&1&1&1&1&1&1&0&0&0&0&0&0&0\\
\end{array}\right)
}
&
{
K_{24}^{\ast} = \left(\begin{array}{cccccccccccccccccccccccc}
1&0&0&0&0&0&0&0&0&0&0&0&0&0&0&0&0&0&0&0&0&0&0&0\\
1&0&0&0&0&0&0&0&0&0&0&0&0&0&0&0&1&0&0&0&0&0&0&0\\
1&0&0&0&1&0&0&0&0&0&0&0&0&0&0&0&0&0&0&0&0&0&0&0\\
1&0&0&0&0&0&0&0&1&0&0&0&0&0&0&0&0&0&0&0&0&0&0&0\\
1&0&1&0&0&0&0&0&0&0&0&0&0&0&0&0&0&0&0&0&0&0&0&0\\
1&1&0&0&0&0&0&0&0&0&0&0&0&0&0&0&0&0&0&0&0&0&0&0\\
1&0&0&0&1&0&0&0&1&0&0&0&1&0&0&0&0&0&0&0&0&0&0&0\\
1&0&1&0&0&0&0&0&1&0&1&0&0&0&0&0&0&0&0&0&0&0&0&0\\
1&1&0&0&0&0&0&0&1&1&0&0&0&0&0&0&0&0&0&0&0&0&0&0\\
1&0&1&0&1&0&1&0&0&0&0&0&0&0&0&0&0&0&0&0&0&0&0&0\\
1&1&0&0&1&1&0&0&0&0&0&0&0&0&0&0&0&0&0&0&0&0&0&0\\
1&1&1&1&0&0&0&0&0&0&0&0&0&0&0&0&0&0&0&0&0&0&0&0\\
0&0&0&0&0&1&1&0&1&1&1&0&1&0&0&0&1&0&0&0&1&0&0&0\\
0&1&1&1&0&0&1&0&1&0&0&0&1&0&0&0&1&0&1&0&0&0&0&0\\
1&0&1&0&1&0&1&0&1&1&0&0&0&0&0&0&1&1&0&0&0&0&0&0\\
0&1&1&0&1&0&1&0&0&1&1&0&1&0&1&0&0&0&0&0&0&0&0&0\\
0&0&1&1&1&1&0&0&1&1&0&0&1&1&0&0&0&0&0&0&0&0&0&0\\
1&1&0&0&1&1&0&0&1&1&1&1&0&0&0&0&0&0&0&0&0&0&0&0\\
1&1&1&1&1&1&1&1&0&0&0&0&0&0&0&0&0&0&0&0&0&0&0&0\\
1&0&1&0&1&0&1&0&1&0&1&0&1&0&1&0&1&0&1&0&1&0&1&0\\
1&1&0&0&1&1&0&0&1&1&0&0&1&1&0&0&1&1&0&0&1&1&0&0\\
1&1&1&1&0&0&0&0&1&1&1&1&0&0&0&0&1&1&1&1&0&0&0&0\\
1&1&1&1&1&1&1&1&0&0&0&0&0&0&0&0&1&1&1&1&1&1&1&1\\
1&1&1&1&1&1&1&1&1&1&1&1&1&1&1&1&0&0&0&0&0&0&0&0\\
\end{array}\right)
} 
& 
{
K_{25}^{\ast} = \left(\begin{array}{ccccccccccccccccccccccccc}
1&0&0&0&0&0&0&0&0&0&0&0&0&0&0&0&0&0&0&0&0&0&0&0&0\\
1&0&0&0&0&0&1&0&0&0&0&0&0&0&0&0&0&0&0&0&0&0&0&0&0\\
1&0&0&0&0&1&0&0&0&0&0&0&0&0&0&0&0&0&0&0&0&0&0&0&0\\
1&0&0&0&1&0&0&0&0&0&0&0&0&0&0&0&0&0&0&0&0&0&0&0&0\\
1&0&1&0&0&0&0&0&0&0&0&0&0&0&0&0&0&0&0&0&0&0&0&0&0\\
1&1&0&0&0&0&0&0&0&0&0&0&0&0&0&0&0&0&0&0&0&0&0&0&0\\
0&1&0&0&0&1&1&0&0&0&0&0&0&1&0&0&0&0&0&0&0&0&0&0&0\\
0&0&1&0&0&1&1&0&0&1&0&0&0&0&0&0&0&0&0&0&0&0&0&0&0\\
0&0&0&1&1&0&0&0&0&1&1&0&0&0&0&0&0&0&0&0&0&0&0&0&0\\
0&0&1&0&1&1&0&0&1&0&0&0&0&0&0&0&0&0&0&0&0&0&0&0&0\\
1&1&1&1&0&0&0&0&0&0&0&0&0&0&0&0&0&0&0&0&0&0&0&0&0\\
0&0&1&0&1&1&0&0&0&0&1&0&0&1&0&1&0&0&0&0&0&0&0&0&0\\
1&1&0&1&0&0&1&0&0&0&1&0&1&0&0&0&0&0&0&0&0&0&0&0&0\\
1&1&1&0&0&1&1&1&0&0&0&0&0&0&0&0&0&0&0&0&0&0&0&0&0\\
0&1&0&1&0&0&1&1&1&0&0&0&0&1&0&1&0&1&0&0&0&0&0&0&0\\
0&0&1&0&1&1&1&1&0&1&0&0&0&0&0&0&0&1&0&0&0&1&0&0&0\\
0&0&0&1&1&0&0&1&1&1&1&0&0&0&0&1&1&0&0&0&0&0&0&0&0\\
1&1&1&1&0&0&0&1&0&0&0&0&1&1&0&0&0&0&0&1&0&0&0&0&0\\
0&1&1&1&0&0&1&1&0&1&1&1&0&0&0&0&0&0&0&0&0&0&0&0&0\\
1&0&0&1&1&1&1&1&1&1&0&1&0&0&0&0&0&0&1&0&0&0&0&0&0\\
1&1&0&1&0&1&0&1&0&1&0&1&0&1&0&1&0&0&0&1&0&1&0&1&0\\
1&1&1&0&0&1&1&0&0&1&1&0&0&1&1&0&0&0&1&0&0&1&1&0&0\\
1&1&1&1&1&0&0&0&0&1&1&1&1&0&0&0&0&0&1&1&1&0&0&0&0\\
1&1&1&1&1&1&1&1&1&0&0&0&0&0&0&0&0&0&1&1&1&1&1&1&1\\
1&1&1&1&1&1&1&1&1&1&1&1&1&1&1&1&1&1&0&0&0&0&0&0&0\\
\end{array}\right)
} 
\end{array}
\\
\begin{array}{cc}
{
\hspace{32mm}K_{26}^{\ast} = \left(
\begin{array}{cccccccccccccccccccccccccc}
1&0&0&0&0&0&0&0&0&0&0&0&0&0&0&0&0&0&0&0&0&0&0&0&0&0\\
1&0&0&0&0&0&0&0&0&1&0&0&0&0&0&0&0&0&0&0&0&0&0&0&0&0\\
1&0&0&0&1&0&0&0&0&0&0&0&0&0&0&0&0&0&0&0&0&0&0&0&0&0\\
1&0&0&1&0&0&0&0&0&0&0&0&0&0&0&0&0&0&0&0&0&0&0&0&0&0\\
1&0&1&0&0&0&0&0&0&0&0&0&0&0&0&0&0&0&0&0&0&0&0&0&0&0\\
1&1&0&0&0&0&0&0&0&0&0&0&0&0&0&0&0&0&0&0&0&0&0&0&0&0\\
1&0&0&1&1&0&0&0&0&0&0&1&0&0&0&0&0&0&0&0&0&0&0&0&0&0\\
1&0&1&0&1&0&0&0&1&0&0&0&0&0&0&0&0&0&0&0&0&0&0&0&0&0\\
1&0&0&0&1&0&0&0&0&1&0&0&0&1&0&0&0&0&0&0&0&0&0&0&0&0\\
0&1&1&1&0&0&0&1&0&0&0&0&0&0&0&0&0&0&0&0&0&0&0&0&0&0\\
1&1&0&1&0&0&1&0&0&0&0&0&0&0&0&0&0&0&0&0&0&0&0&0&0&0\\
0&0&0&1&0&0&0&1&1&1&0&1&0&0&0&0&0&0&0&0&1&0&0&0&0&0\\
0&1&0&0&1&0&0&0&1&0&1&0&0&1&0&1&0&0&0&0&0&0&0&0&0&0\\
1&1&0&1&0&0&0&1&0&1&1&0&0&0&0&0&0&0&0&0&0&0&0&0&0&0\\
1&1&1&1&1&1&0&0&0&0&0&0&0&0&0&0&0&0&0&0&0&0&0&0&0&0\\
1&0&0&1&1&0&1&1&0&0&0&1&0&1&1&0&0&0&0&0&0&0&0&0&0&0\\
0&1&1&0&0&0&1&1&0&1&1&0&0&1&0&0&0&0&1&0&0&0&0&0&0&0\\
1&1&1&0&1&0&1&1&1&0&0&0&1&0&0&0&0&0&0&0&0&0&0&0&0&0\\
0&0&0&1&0&1&1&1&0&0&1&0&1&0&0&0&0&1&1&0&1&0&1&0&0&0\\
0&1&1&0&1&1&0&1&1&0&0&0&0&1&0&0&0&0&1&0&1&0&0&0&1&0\\
1&0&1&0&1&0&1&0&0&1&1&1&1&0&0&1&0&1&0&0&0&0&0&0&0&0\\
1&1&0&0&1&1&0&1&1&0&1&0&1&0&1&0&0&1&0&0&1&1&0&0&0&0\\
1&1&1&1&0&0&0&1&1&1&0&0&1&1&0&0&0&1&1&1&0&0&0&0&0&0\\
1&1&1&1&1&1&1&0&0&0&0&0&1&1&1&1&1&0&0&0&0&0&0&0&0&0\\
1&1&1&1&1&1&1&1&1&1&1&1&0&0&0&0&0&0&0&0&0&0&0&0&1&1\\
1&1&1&1&1&1&1&1&1&1&1&1&1&1&1&1&1&1&1&1&1&1&1&1&0&0\\
\end{array}\right)
}
&
{
K_{27}^{\ast} = \left(\begin{array}{ccccccccccccccccccccccccccc}
1&0&0&0&0&0&0&0&0&0&0&0&0&0&0&0&0&0&0&0&0&0&0&0&0&0&0\\
1&0&0&0&0&0&0&1&0&0&0&0&0&0&0&0&0&0&0&0&0&0&0&0&0&0&0\\
1&0&0&0&0&0&1&0&0&0&0&0&0&0&0&0&0&0&0&0&0&0&0&0&0&0&0\\
1&0&0&1&0&0&0&0&0&0&0&0&0&0&0&0&0&0&0&0&0&0&0&0&0&0&0\\
1&0&1&0&0&0&0&0&0&0&0&0&0&0&0&0&0&0&0&0&0&0&0&0&0&0&0\\
1&1&0&0&0&0&0&0&0&0&0&0&0&0&0&0&0&0&0&0&0&0&0&0&0&0&0\\
0&1&1&1&0&1&0&0&0&0&0&0&0&0&0&0&0&0&0&0&0&0&0&0&0&0&0\\
0&0&0&1&0&1&0&0&1&0&0&1&0&0&0&0&0&0&0&0&0&0&0&0&0&0&0\\
0&0&1&1&0&0&1&0&0&1&0&0&0&0&0&0&0&0&0&0&0&0&0&0&0&0&0\\
1&0&0&1&0&1&0&0&0&0&0&0&0&1&0&0&0&0&0&0&0&0&0&0&0&0&0\\
1&1&0&1&1&0&0&0&0&0&0&0&0&0&0&0&0&0&0&0&0&0&0&0&0&0&0\\
1&0&0&0&0&0&1&1&0&1&0&1&0&0&1&0&0&0&0&0&0&0&0&0&0&0&0\\
1&1&0&0&0&1&0&0&1&1&1&0&0&0&0&0&0&0&0&0&0&0&0&0&0&0&0\\
0&0&0&0&0&0&1&0&1&0&1&1&1&0&0&0&0&0&1&0&0&0&0&0&0&0&0\\
0&1&1&1&0&0&1&1&1&0&0&0&0&0&0&0&0&0&0&0&0&0&0&0&0&0&0\\
0&0&0&0&1&0&1&0&1&1&1&1&0&1&0&0&0&0&0&0&0&0&1&0&0&0&0\\
0&0&0&1&0&1&1&1&1&0&0&1&0&0&0&0&0&0&1&0&0&0&0&0&1&0&0\\
1&0&0&1&0&1&0&1&1&0&1&0&0&1&0&0&0&1&0&0&0&0&0&0&0&0&0\\
1&1&1&1&1&1&0&0&1&0&0&0&1&0&0&0&0&0&0&0&0&0&0&0&0&0&0\\
1&1&0&0&1&1&1&0&0&0&1&0&1&1&1&0&0&0&0&0&0&1&0&0&0&0&0\\
1&0&1&0&0&1&1&0&1&1&0&1&0&1&1&0&1&0&0&0&0&0&0&0&0&0&0\\
0&0&1&1&1&1&1&1&0&1&0&0&1&1&0&1&0&0&0&0&0&0&0&0&0&0&0\\
1&1&0&0&1&0&1&1&0&1&1&0&1&0&0&1&1&0&1&0&1&0&0&0&0&0&0\\
1&1&1&1&0&0&1&1&1&0&0&0&1&1&1&0&0&0&1&1&0&0&0&0&0&0&0\\
1&1&1&1&1&1&0&0&0&0&0&0&1&1&1&1&1&1&0&0&0&0&0&0&1&0&1\\
1&1&1&1&1&1&1&1&1&1&1&1&0&0&0&0&0&0&0&0&0&0&0&0&1&1&0\\
1&1&1&1&1&1&1&1&1&1&1&1&1&1&1&1&1&1&1&1&1&1&1&1&0&0&0\\
\end{array}\right)
} 
\end{array}
$
}}
\caption{Polarization kernels with the best rate of polarization}
\label{fBestKernels}
\end{figure*}

\begin{figure*}
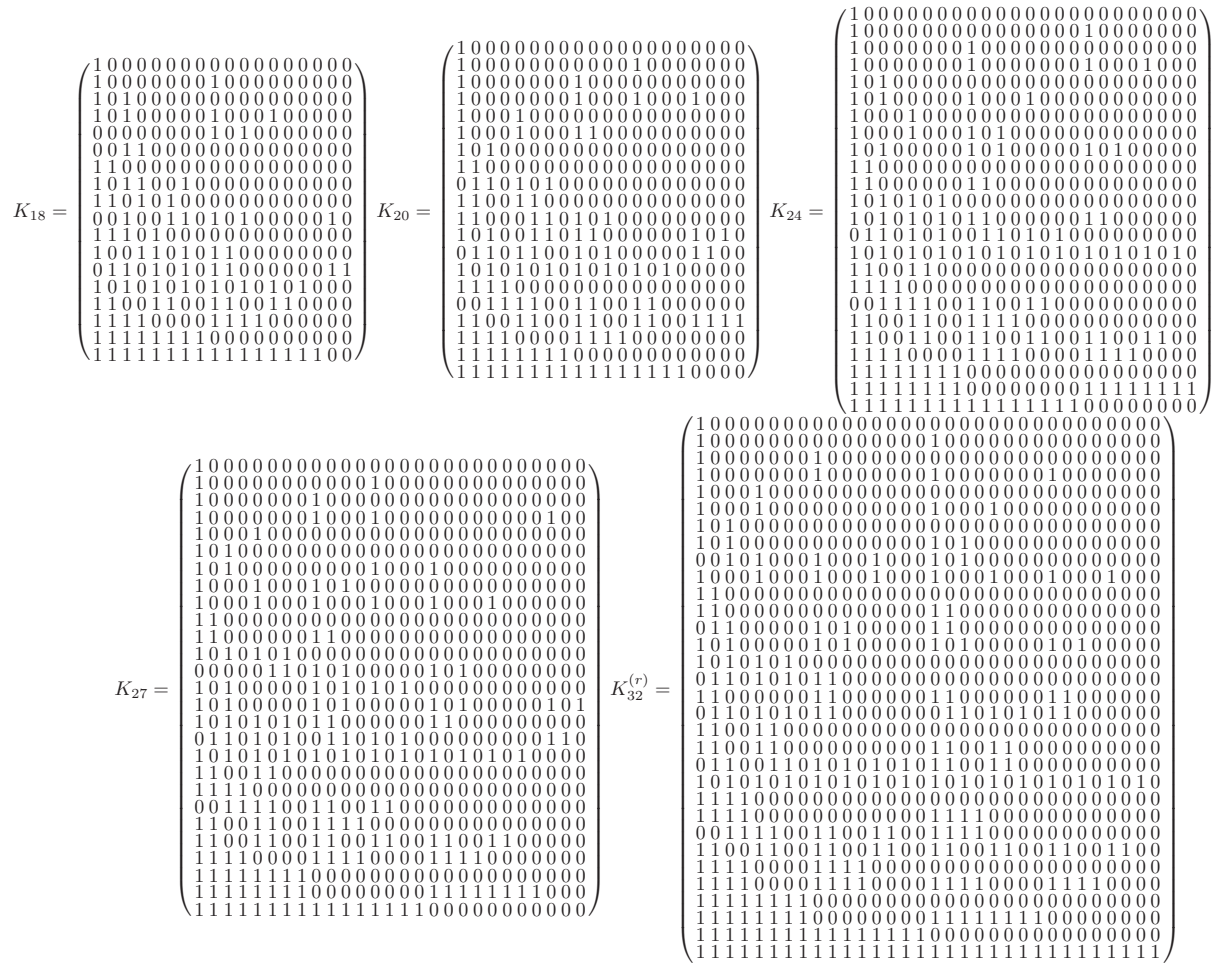

\centering
\scalebox{0.75}{
\parbox{1.2\textwidth}
{ 
$
\arraycolsep=1.2pt\def\arraystretch{0.6}
\begin{array}{ccc}
K_{18} = \left(\begin{array}{cccccccccccccccccc}
1&0&0&0&0&0&0&0&0&0&0&0&0&0&0&0&0&0\\
1&0&0&0&0&0&0&0&1&0&0&0&0&0&0&0&0&0\\
1&0&1&0&0&0&0&0&0&0&0&0&0&0&0&0&0&0\\
1&0&1&0&0&0&0&0&1&0&0&0&1&0&0&0&0&0\\
0&0&0&0&0&0&0&0&1&0&1&0&0&0&0&0&0&0\\
0&0&1&1&0&0&0&0&0&0&0&0&0&0&0&0&0&0\\
1&1&0&0&0&0&0&0&0&0&0&0&0&0&0&0&0&0\\
1&0&1&1&0&0&1&0&0&0&0&0&0&0&0&0&0&0\\
1&1&0&1&0&1&0&0&0&0&0&0&0&0&0&0&0&0\\
0&0&1&0&0&1&1&0&1&0&1&0&0&0&0&0&1&0\\
1&1&1&0&1&0&0&0&0&0&0&0&0&0&0&0&0&0\\
1&0&0&1&1&0&1&0&1&1&0&0&0&0&0&0&0&0\\
0&1&1&0&1&0&1&0&1&1&0&0&0&0&0&0&1&1\\
1&0&1&0&1&0&1&0&1&0&1&0&1&0&1&0&0&0\\
1&1&0&0&1&1&0&0&1&1&0&0&1&1&0&0&0&0\\
1&1&1&1&0&0&0&0&1&1&1&1&0&0&0&0&0&0\\
1&1&1&1&1&1&1&1&0&0&0&0&0&0&0&0&0&0\\
1&1&1&1&1&1&1&1&1&1&1&1&1&1&1&1&0&0\\
\end{array}\right)
&
{
K_{20} = \left(\begin{array}{cccccccccccccccccccc}
1&0&0&0&0&0&0&0&0&0&0&0&0&0&0&0&0&0&0&0\\
1&0&0&0&0&0&0&0&0&0&0&0&1&0&0&0&0&0&0&0\\
1&0&0&0&0&0&0&0&1&0&0&0&0&0&0&0&0&0&0&0\\
1&0&0&0&0&0&0&0&1&0&0&0&1&0&0&0&1&0&0&0\\
1&0&0&0&1&0&0&0&0&0&0&0&0&0&0&0&0&0&0&0\\
1&0&0&0&1&0&0&0&1&1&0&0&0&0&0&0&0&0&0&0\\
1&0&1&0&0&0&0&0&0&0&0&0&0&0&0&0&0&0&0&0\\
1&1&0&0&0&0&0&0&0&0&0&0&0&0&0&0&0&0&0&0\\
0&1&1&0&1&0&1&0&0&0&0&0&0&0&0&0&0&0&0&0\\
1&1&0&0&1&1&0&0&0&0&0&0&0&0&0&0&0&0&0&0\\
1&1&0&0&0&1&1&0&1&0&1&0&0&0&0&0&0&0&0&0\\
1&0&1&0&0&1&1&0&1&1&0&0&0&0&0&0&1&0&1&0\\
0&1&1&0&1&1&0&0&1&0&1&0&0&0&0&0&1&1&0&0\\
1&0&1&0&1&0&1&0&1&0&1&0&1&0&1&0&0&0&0&0\\
1&1&1&1&0&0&0&0&0&0&0&0&0&0&0&0&0&0&0&0\\
0&0&1&1&1&1&0&0&1&1&0&0&1&1&0&0&0&0&0&0\\
1&1&0&0&1&1&0&0&1&1&0&0&1&1&0&0&1&1&1&1\\
1&1&1&1&0&0&0&0&1&1&1&1&0&0&0&0&0&0&0&0\\
1&1&1&1&1&1&1&1&0&0&0&0&0&0&0&0&0&0&0&0\\
1&1&1&1&1&1&1&1&1&1&1&1&1&1&1&1&0&0&0&0\\
\end{array}\right)
}
&
{
K_{24} = \left(
\begin{array}{cccccccccccccccccccccccc}
1&0&0&0&0&0&0&0&0&0&0&0&0&0&0&0&0&0&0&0&0&0&0&0\\
1&0&0&0&0&0&0&0&0&0&0&0&0&0&0&0&1&0&0&0&0&0&0&0\\
1&0&0&0&0&0&0&0&1&0&0&0&0&0&0&0&0&0&0&0&0&0&0&0\\
1&0&0&0&0&0&0&0&1&0&0&0&0&0&0&0&1&0&0&0&1&0&0&0\\
1&0&1&0&0&0&0&0&0&0&0&0&0&0&0&0&0&0&0&0&0&0&0&0\\
1&0&1&0&0&0&0&0&1&0&0&0&1&0&0&0&0&0&0&0&0&0&0&0\\
1&0&0&0&1&0&0&0&0&0&0&0&0&0&0&0&0&0&0&0&0&0&0&0\\
1&0&0&0&1&0&0&0&1&0&1&0&0&0&0&0&0&0&0&0&0&0&0&0\\
1&0&1&0&0&0&0&0&1&0&1&0&0&0&0&0&1&0&1&0&0&0&0&0\\
1&1&0&0&0&0&0&0&0&0&0&0&0&0&0&0&0&0&0&0&0&0&0&0\\
1&1&0&0&0&0&0&0&1&1&0&0&0&0&0&0&0&0&0&0&0&0&0&0\\
1&0&1&0&1&0&1&0&0&0&0&0&0&0&0&0&0&0&0&0&0&0&0&0\\
1&0&1&0&1&0&1&0&1&1&0&0&0&0&0&0&1&1&0&0&0&0&0&0\\
0&1&1&0&1&0&1&0&0&1&1&0&1&0&1&0&0&0&0&0&0&0&0&0\\
1&0&1&0&1&0&1&0&1&0&1&0&1&0&1&0&1&0&1&0&1&0&1&0\\
1&1&0&0&1&1&0&0&0&0&0&0&0&0&0&0&0&0&0&0&0&0&0&0\\
1&1&1&1&0&0&0&0&0&0&0&0&0&0&0&0&0&0&0&0&0&0&0&0\\
0&0&1&1&1&1&0&0&1&1&0&0&1&1&0&0&0&0&0&0&0&0&0&0\\
1&1&0&0&1&1&0&0&1&1&1&1&0&0&0&0&0&0&0&0&0&0&0&0\\
1&1&0&0&1&1&0&0&1&1&0&0&1&1&0&0&1&1&0&0&1&1&0&0\\
1&1&1&1&0&0&0&0&1&1&1&1&0&0&0&0&1&1&1&1&0&0&0&0\\
1&1&1&1&1&1&1&1&0&0&0&0&0&0&0&0&0&0&0&0&0&0&0&0\\
1&1&1&1&1&1&1&1&0&0&0&0&0&0&0&0&1&1&1&1&1&1&1&1\\
1&1&1&1&1&1&1&1&1&1&1&1&1&1&1&1&0&0&0&0&0&0&0&0\\
\end{array}\right)
} 
\end{array}
\\
\begin{array}{cc}
{
\hspace{18mm}K_{27} = \left(
\begin{array}{ccccccccccccccccccccccccccc}
1&0&0&0&0&0&0&0&0&0&0&0&0&0&0&0&0&0&0&0&0&0&0&0&0&0&0\\
1&0&0&0&0&0&0&0&0&0&0&0&1&0&0&0&0&0&0&0&0&0&0&0&0&0&0\\
1&0&0&0&0&0&0&0&1&0&0&0&0&0&0&0&0&0&0&0&0&0&0&0&0&0&0\\
1&0&0&0&0&0&0&0&1&0&0&0&1&0&0&0&0&0&0&0&0&0&0&0&1&0&0\\
1&0&0&0&1&0&0&0&0&0&0&0&0&0&0&0&0&0&0&0&0&0&0&0&0&0&0\\
1&0&1&0&0&0&0&0&0&0&0&0&0&0&0&0&0&0&0&0&0&0&0&0&0&0&0\\
1&0&1&0&0&0&0&0&0&0&0&0&1&0&0&0&1&0&0&0&0&0&0&0&0&0&0\\
1&0&0&0&1&0&0&0&1&0&1&0&0&0&0&0&0&0&0&0&0&0&0&0&0&0&0\\
1&0&0&0&1&0&0&0&1&0&0&0&1&0&0&0&1&0&0&0&1&0&0&0&0&0&0\\
1&1&0&0&0&0&0&0&0&0&0&0&0&0&0&0&0&0&0&0&0&0&0&0&0&0&0\\
1&1&0&0&0&0&0&0&1&1&0&0&0&0&0&0&0&0&0&0&0&0&0&0&0&0&0\\
1&0&1&0&1&0&1&0&0&0&0&0&0&0&0&0&0&0&0&0&0&0&0&0&0&0&0\\
0&0&0&0&0&1&1&0&1&0&1&0&0&0&0&0&1&0&1&0&0&0&0&0&0&0&0\\
1&0&1&0&0&0&0&0&1&0&1&0&1&0&1&0&0&0&0&0&0&0&0&0&0&0&0\\
1&0&1&0&0&0&0&0&1&0&1&0&0&0&0&0&1&0&1&0&0&0&0&0&1&0&1\\
1&0&1&0&1&0&1&0&1&1&0&0&0&0&0&0&1&1&0&0&0&0&0&0&0&0&0\\
0&1&1&0&1&0&1&0&0&1&1&0&1&0&1&0&0&0&0&0&0&0&0&0&1&1&0\\
1&0&1&0&1&0&1&0&1&0&1&0&1&0&1&0&1&0&1&0&1&0&1&0&0&0&0\\
1&1&0&0&1&1&0&0&0&0&0&0&0&0&0&0&0&0&0&0&0&0&0&0&0&0&0\\
1&1&1&1&0&0&0&0&0&0&0&0&0&0&0&0&0&0&0&0&0&0&0&0&0&0&0\\
0&0&1&1&1&1&0&0&1&1&0&0&1&1&0&0&0&0&0&0&0&0&0&0&0&0&0\\
1&1&0&0&1&1&0&0&1&1&1&1&0&0&0&0&0&0&0&0&0&0&0&0&0&0&0\\
1&1&0&0&1&1&0&0&1&1&0&0&1&1&0&0&1&1&0&0&1&1&0&0&0&0&0\\
1&1&1&1&0&0&0&0&1&1&1&1&0&0&0&0&1&1&1&1&0&0&0&0&0&0&0\\
1&1&1&1&1&1&1&1&0&0&0&0&0&0&0&0&0&0&0&0&0&0&0&0&0&0&0\\
1&1&1&1&1&1&1&1&0&0&0&0&0&0&0&0&1&1&1&1&1&1&1&1&0&0&0\\
1&1&1&1&1&1&1&1&1&1&1&1&1&1&1&1&0&0&0&0&0&0&0&0&0&0&0\\
\end{array}\right)
}
&
{
K_{32}^{(r)} = \left(\begin{array}{cccccccccccccccccccccccccccccccc}
1&0&0&0&0&0&0&0&0&0&0&0&0&0&0&0&0&0&0&0&0&0&0&0&0&0&0&0&0&0&0&0\\
1&0&0&0&0&0&0&0&0&0&0&0&0&0&0&0&1&0&0&0&0&0&0&0&0&0&0&0&0&0&0&0\\
1&0&0&0&0&0&0&0&1&0&0&0&0&0&0&0&0&0&0&0&0&0&0&0&0&0&0&0&0&0&0&0\\
1&0&0&0&0&0&0&0&1&0&0&0&0&0&0&0&1&0&0&0&0&0&0&0&1&0&0&0&0&0&0&0\\
1&0&0&0&1&0&0&0&0&0&0&0&0&0&0&0&0&0&0&0&0&0&0&0&0&0&0&0&0&0&0&0\\
1&0&0&0&1&0&0&0&0&0&0&0&0&0&0&0&1&0&0&0&1&0&0&0&0&0&0&0&0&0&0&0\\
1&0&1&0&0&0&0&0&0&0&0&0&0&0&0&0&0&0&0&0&0&0&0&0&0&0&0&0&0&0&0&0\\
1&0&1&0&0&0&0&0&0&0&0&0&0&0&0&0&1&0&1&0&0&0&0&0&0&0&0&0&0&0&0&0\\
0&0&1&0&1&0&0&0&1&0&0&0&1&0&0&0&1&0&1&0&0&0&0&0&0&0&0&0&0&0&0&0\\
1&0&0&0&1&0&0&0&1&0&0&0&1&0&0&0&1&0&0&0&1&0&0&0&1&0&0&0&1&0&0&0\\
1&1&0&0&0&0&0&0&0&0&0&0&0&0&0&0&0&0&0&0&0&0&0&0&0&0&0&0&0&0&0&0\\
1&1&0&0&0&0&0&0&0&0&0&0&0&0&0&0&1&1&0&0&0&0&0&0&0&0&0&0&0&0&0&0\\
0&1&1&0&0&0&0&0&1&0&1&0&0&0&0&0&1&1&0&0&0&0&0&0&0&0&0&0&0&0&0&0\\
1&0&1&0&0&0&0&0&1&0&1&0&0&0&0&0&1&0&1&0&0&0&0&0&1&0&1&0&0&0&0&0\\
1&0&1&0&1&0&1&0&0&0&0&0&0&0&0&0&0&0&0&0&0&0&0&0&0&0&0&0&0&0&0&0\\
0&1&1&0&1&0&1&0&1&1&0&0&0&0&0&0&0&0&0&0&0&0&0&0&0&0&0&0&0&0&0&0\\
1&1&0&0&0&0&0&0&1&1&0&0&0&0&0&0&1&1&0&0&0&0&0&0&1&1&0&0&0&0&0&0\\
0&1&1&0&1&0&1&0&1&1&0&0&0&0&0&0&0&1&1&0&1&0&1&0&1&1&0&0&0&0&0&0\\
1&1&0&0&1&1&0&0&0&0&0&0&0&0&0&0&0&0&0&0&0&0&0&0&0&0&0&0&0&0&0&0\\
1&1&0&0&1&1&0&0&0&0&0&0&0&0&0&0&1&1&0&0&1&1&0&0&0&0&0&0&0&0&0&0\\
0&1&1&0&0&1&1&0&1&0&1&0&1&0&1&0&1&1&0&0&1&1&0&0&0&0&0&0&0&0&0&0\\
1&0&1&0&1&0&1&0&1&0&1&0&1&0&1&0&1&0&1&0&1&0&1&0&1&0&1&0&1&0&1&0\\
1&1&1&1&0&0&0&0&0&0&0&0&0&0&0&0&0&0&0&0&0&0&0&0&0&0&0&0&0&0&0&0\\
1&1&1&1&0&0&0&0&0&0&0&0&0&0&0&0&1&1&1&1&0&0&0&0&0&0&0&0&0&0&0&0\\
0&0&1&1&1&1&0&0&1&1&0&0&1&1&0&0&1&1&1&1&0&0&0&0&0&0&0&0&0&0&0&0\\
1&1&0&0&1&1&0&0&1&1&0&0&1&1&0&0&1&1&0&0&1&1&0&0&1&1&0&0&1&1&0&0\\
1&1&1&1&0&0&0&0&1&1&1&1&0&0&0&0&0&0&0&0&0&0&0&0&0&0&0&0&0&0&0&0\\
1&1&1&1&0&0&0&0&1&1&1&1&0&0&0&0&1&1&1&1&0&0&0&0&1&1&1&1&0&0&0&0\\
1&1&1&1&1&1&1&1&0&0&0&0&0&0&0&0&0&0&0&0&0&0&0&0&0&0&0&0&0&0&0&0\\
1&1&1&1&1&1&1&1&0&0&0&0&0&0&0&0&1&1&1&1&1&1&1&1&0&0&0&0&0&0&0&0\\
1&1&1&1&1&1&1&1&1&1&1&1&1&1&1&1&0&0&0&0&0&0&0&0&0&0&0&0&0&0&0&0\\
1&1&1&1&1&1&1&1&1&1&1&1&1&1&1&1&1&1&1&1&1&1&1&1&1&1&1&1&1&1&1&1\\

\end{array}\right)
} 
\end{array}
$
}}
\caption{Polarization kernels which admit low complexity processing}
\label{fSimpleKernels}
\end{figure*}

Fig. \ref{fBestKernels} presents  polarization kernels with the best rate of polarization, obtained in Section \ref{sMaximization}. Fig. \ref{fSimpleKernels} presents  polarization kernels which admit low complexity processing, obtained in Section \ref{sComplexity}.